\documentclass[11pt,a4paper]{article}
\pdfoutput=1

\usepackage[%
bibstyle=numeric,%
style=numeric-comp,%
sorting=none,%
sortcites=true,%
maxnames=3,%
abbreviate=true,%
backend=bibtex,%
]{biblatex}

\bibliography{GaugeBetheRef}
\usepackage[]{Packages}

\usepackage{Definitions}

\definecolor{Red}{rgb}{1,0,0}
\definecolor{Blue}{rgb}{0,0,1}
\definecolor{Green}{rgb}{0,1,0}
\definecolor{DarkGreen}{rgb}{0,.64,0}
\definecolor{chartreuse}{rgb}{.49,.98,0}
\definecolor{amethyst}{rgb}{0.59375,0.398438,0.792969}
\definecolor{brownrust}{rgb}{0.6875, 0.316406, 0.242188}
\definecolor{FreshEggplant}{rgb}{0.59375, 0., 0.414063}
\definecolor{BurntOrange}{rgb}{0.792969,0.332031,0}
\definecolor{FreshEggplant}{rgb}{0.59375, 0., 0.414063}
\definecolor{Salmon}{rgb}{1,0.55,0.41}
\definecolor{salmon}{rgb}{0.996094,0.507813,0.410156}
 \definecolor{FrenchRose}{rgb}{0.96875, 0.292969, 0.5625}
\definecolor{Cabaret}{rgb}{0.808594, 0.242188, 0.46875}
\definecolor{Shamrock}{rgb}{0.242188, 0.808594, 0.582031}
\definecolor{RobinsEggBlue}{rgb}{0., 0.792969, 0.792969}
\definecolor{GuardsmanRed}{rgb}{0.792969, 0., 0.}
\definecolor{Sapphire}{rgb}{0.183594, 0.328125, 0.621094}
\definecolor{Sorbus}{rgb}{0.996094, 0.429688, 0.0273438}

\def\SHCommentColor{Sorbus}

\newcommand{\mshg}[1]{{\color{\SHCommentColor}{#1}}}

\def\muchlessthan{<\hskip-.07in <}
\def\ll{_}
\def\uu{^}
\def\sqd{^2}
\def\O{\Omega}

\def\t{\tau}
\def\phd{\dot{\phi}}
\def\r{\rho}
\def\d{\delta}
\def\s{\sigma}
\def\m{\mu}
\def\n{\nu}
\def\pp{\partial}
\def\lrdd{\bigg (}
\def\rrdd{\bigg )}
\def\lsqq{\bigg [}
\def\rsqq{\bigg ]}
\newcommand*{\rws}{\ensuremath{R_{\textsc{ws}}}}
\def\eng{{\bf E}}
\def\bp{{\bf P}}
\newcommand{\rr}[1]{(\ref{#1})}

\newcommand{\nn}{\nonumber}
\newcommand{\een}[1]{\label{#1} \eee}

\def\xxx{\nn\eee\bbb}
\def\xxn{\eee\bbb}

\def\apr{\alpha^\prime{}}
\def\llsk{{\hskip .25in}}
\def\pr{{}^\prime}
\def\ffk#1#2{\frac {{#1}} {{#2}}}
\def\ww{\wedge}
\def\kko{\ ,}
\def\ppo{\ .}
\def\cl{{\cal L}}
\def\ch{{\cal H}}

\allowdisplaybreaks[1]

\preprint{\texttt{CERN-PH-TH/2012-285}\\\texttt{IPMU12--0194}}

\newcommand{\OfficialTitle}{BPS States in the Duality Web\\ of the Omega deformation}

\title{\vspace{2cm}
  {\huge   \textbf{\sffamily\OfficialTitle}}
}

\hypersetup{pdfauthor={Simeon Hellerman and Domenico Orlando and Susanne Reffert},pdftitle={\OfficialTitle}}

\author{
  \begin{minipage}{.8\linewidth}
    \vspace{1cm}
    \begin{center}
      {\small \textbf{Simeon Hellerman}\( {}^\flat \), \textbf{Domenico Orlando}\( {}^\# \) and \textbf{Susanne Reffert}\( {}^\# \) }
    \end{center}
    \vspace{1cm}
    \begin{minipage}{\linewidth}\centering
      {\itshape\footnotesize 
        \begin{itemize}
        \item[\( {}^\flat \)] Kavli Institute for the Physics and Mathematics of the Universe (WPI),\\ The University of Tokyo (ToDIAS), \\ Kashiwa, Chiba 277--8568, Japan.
        \item[\( {}^\# \)] Theory Group, Physics Department, \\ Organisation européenne pour la recherche nucléaire (CERN) \\ CH-1211 Geneva 23, Switzerland
        \end{itemize}
      }
    \end{minipage}
  \end{minipage}
}

\date{}

\begin{document}

\setstretch{1.1}

\numberwithin{equation}{section}

\begin{titlepage}

  \maketitle

  \thispagestyle{empty}

  \vfill
  \abstract{
  In this note, we study different limits of an $\Omega$--deformed \( (2,0) \) six-dimensional gauge theory realized in an M--theory fluxtrap background. Via a chain of dualities, we connect the $\O$--deformed \acs{sym} to a new four-dimensional gauge theory which we refer to as the reciprocal gauge theory. This theory has several properties in common with Liouville field theory, such as its gauge coupling $b\sqd =  \e\ll 2 / \e\ll 1$, and its behavior under S--duality. Finally, we realize the \acs{bps} states on the \acs{sym} side of the \acs{agt} correspondence and follow them along the chain of dualities. 
 In the fluxtrap frame, we are dealing with two distinct types of states localized in different radial positions, while in the reciprocal frame, we find single states carrying both charges localized in one place which appear to be perturbatively stable.  Our microscopic picture of the small-$b$ limit exhibits semiclassically \textsc{bps} bound states, which are not visible at the level of the partition function.
}
\vfill

\end{titlepage}

\section{Introduction}

\def\O{\Omega}

The \ac{agt} correspondence~\cite{Alday:2009aq} is defined
in terms of the $\O$--deformation~\cite{Lossev:1997bz,Moore:1997dj,Nekrasov:2002qd,Nekrasov:2003rj} of four-dimensional supersymmetric
gauge theory.  %
Starting from a gauge theory with $\mathcal{N} = 4$ (or \( \mathcal{N} =2 \)) supersymmetry on a manifold ${\bf M}\ll 4$
with $U(1) \times U(1)$ isometries, there is a particular way of deforming the
theory with respect to those isometries that preserves some supersymmetry.
The
resulting theory has some supersymmetrically protected quantities -- in particular,
the so-called instanton partition function. %
For certain particularly simple ${\bf M}\ll 4$, there is a beautiful connection with
two-dimensional field theory.  In particular when ${\bf M}\ll 4 = S\uu 4$, there
is an equality between the
instanton partition function of various $\mathcal{N} = 2$ and $\mathcal{N} = 4$
theories on the one hand, and amplitudes in two-dimensional Liouville or Toda theories on the other hand.  This equality is known as the \ac{agt} correspondence.

The origin of the two-dimensional theory has not been fully understood.  However various 
aspects of the duality have suggested a connection via  a lift of the $\mathcal{N} = 4$
or $\mathcal{N} = 2$ theory to the six-dimensional quantum field theory with $(2,0)$ superconformal
symmetry, compactified on some Riemann surface $\Sigma$, where the curve $\Sigma$
is related to the Seiberg--Witten curve of the 4-dimensional gauge theory.  In this
context, certain things are very natural: For instance, partition functions of gauge
theories as a function of their coupling and mass transform under dualities in a way
that follows the modular transformations of punctured Riemann surfaces, with
the masses and couplings of the gauge theory parametrizing the Teichm\"uller
space of the punctured Riemann surface.
Nonetheless the origin of the detailed dynamics of Liouville/Toda theory remains
somewhat obscure.  

Another simple case to consider is the case where ${\bf M}\ll 4 = \IC\uu 2$.  Here, there
is a heuristic sense in which one feels the corresponding two-dimensional theory
ought to be the ``chiral half'' of Liouville theory according to some definition.  As we shall
show in this paper, this is too na\"ive, and there is no sense in which the partition function
on the $\O$--deformed $\IC\uu 2$ corresponds to the partition function of a two-dimensional quantum field theory at all:  the gauge theory has an infinite volume region orthogonal to the
two translationally invariant compact directions, with
momentum continua and adjustable vacuum expectation values
at infinity in these additional directions.
Nonetheless this theory has
characteristic properties that parallel those of the two-dimensional
Liouville/Toda theory.  In particular it is equipped with a certain semiclassical limit, where the
ratio of the parameters $\e\ll {1,2}$ describing the $\O$-deformation, goes to
zero.

In recent work, the authors constructed a background of string theory whose low-energy
dynamics describes the $\O$--deformed four-dimensional gauge theory with
$\mathcal{N} = 2$ supersymmetry~\cite{Orlando:2010uu,Orlando:2010aj,Hellerman:2011mv,Hellerman:2012zf}.  We subsequently lifted the construction
to $11$-dimensional M--theory, realizing the gauge theory as the dynamics of an
M5--brane on a particular curve~$\Sigma$, deformed by the presence of flux and metric
curvature.  In the present article, we shall describe the
corresponding deformation of the $\mathcal{N} = 4$ gauge theory, the $\O$--deformation of the string
background, and its lift to M--theory, with generic deformation parameters $\e\ll{1,2}$.  We shall then reduce the theory on the $U(1) \times U(1)$ isometry orbits to obtain a new solution
of \tIIB string theory, where the gauge dynamics is realized on a D3--brane with a gauge
coupling proportional to $\e\ll 2 / \e\ll 1$.   This theory is noncompact in two of its
four dimensions, and the dynamics are four- rather than two-dimensional.

The paper is organized as follows.  In Section~\ref{sec:from-m-theory} we describe the
$\mathcal{N} = 4$ version of our earlier construction, lift it to M--theory, and then
reduce on the $U(1) \times U(1)$ isometry orbits to a solution of \tIIB 
string theory that we refer to as the \emph{reciprocal} duality frame.  In Section~\ref{sec:weekly-coupl-theory} we follow the brane dynamics of the D3--branes on which the original
$\mathcal{N} = 4$ gauge theory was realized, through their transmutation into M5--branes
supporting a full (2,0) dynamics, back into D3--branes with a different gauge coupling and
background metric, whose dynamics generate a four-dimensional gauge theory. %
In particular we discuss its behavior under S--duality, which we find to be parallel to the strong/weak coupling duality of Liouville/Toda theory realized as the transformation $b\to 1/b$. 
In Section~\ref{sec:BPSandDOZZ} we
compute the spectrum
of \ac{bps} states of the original $\mathcal{N} = 4$ gauge theory, tracing them through the duality web
to their incarnation as \D3--branes of finite volume
in the reciprocal frame.  %
In Section~\ref{sec:conclusions} we briefly present conclusions and outlook on further research. %
In Appendix~\ref{sec:supersymmetry}, we discuss the supersymmetries preserved in the bulk of each duality frame.

\section{Chain of dualities -- the bulk}
\label{sec:from-m-theory}

In this note, we study different limits of the six-dimensional $(2,0)$ theory in the $\O$--background via a chain of dualities starting from a Melvin or fluxbrane background. %
Having started from identifications on flat space, the theory lives on \( \setR^4_{\Omega} \times T^2 \), where \( \setR^4_{\Omega}  \) is the product of two cigars. The compactification on \( T^2 \) gives by construction the \( \Omega \)--deformed \( \mathcal{N} = 4 \) \textsc{sym};  the only other directions on which we can reduce the six-dimensional theory are the two angular directions, which results in a new four-dimensional theory. This theory exhibits properties reminiscent of the two-dimensional Liouville theory in the \ac{agt} correspondence, such as its gauge coupling \( b = \epsilon_2 / \epsilon_1 \) and its behavior under S--duality, thus constituting an important step towards a direct construction of Liouville theory from a string theory setting.

The chain of dualities which we will explain in detail in the following is summarized in Table~\ref{tab:duality-chain}. We start from a fluxbrane background in \tIIB and perform two T--dualities to arrive again at a \tIIB theory, but in a fluxtrap background. After another T-duality to \tIIA and a lift to M--theory, we have reached the deformed M--theory background in which the $(2,0)$ theory lives. From here, reduction in the two angular directions brings us to the  \emph{reciprocal background}, a \tIIB theory with flux and a deformed metric background and dilaton gradient, and additionally a \D5/\NS5 brane at the center of the geometry.

\paragraph{The Fluxbrane.}

\newcommand*{\freccia}[1]{%
  \multicolumn{3}{c}{%
   \hspace*{-.1\textwidth}
    \begin{tikzpicture}
      \node [single arrow, shape border uses incircle,inner sep=1pt, draw, shape border rotate=-90, minimum height=4em, rotate=0] at (0,0) {m};
      \node [style={fill=white,inner sep=2pt} ] at (0,0) {\emph{#1}};
    \end{tikzpicture}
 }
}

\begin{table}
  \centering
  \begin{tabular}{m{.35\textwidth}m{.08\textwidth}m{.48\textwidth}}
    \toprule
    Bulk & Probe & Gauge Theory \\ \midrule
    \tIIB in Melvin space & \D5 & six-dimensional gauge theory with Wilson line boundary conditions in two directions \\[.4em]
    \freccia{T--duality in  $\tilde u_1$  and $ \tilde u_2$} \\  
    \tIIB in complex fluxtrap & \D3 & \( \Omega \)--deformed \( \mathcal{N} = 4 \) SYM \\[.4em]
    \freccia{T--duality in \( \tilde x_6 \) and lift} \\ 
    M--theory fluxtrap & \M5 & \( (2,0) \) six-dimensional theory \\[.4em]
    \freccia{reduction in \( \sigma_1 \) and \( \sigma_2 \)} \\ %
    \tIIB in deformed \D5/\NS5 (\acs{rfold}) & \D3 & \acl{nxw} \\
    \bottomrule
  \end{tabular}
  \caption{The chain of dualities among the different string frames and the corresponding effective gauge theories.}
  \label{tab:duality-chain}
\end{table}

We start out from Euclidean flat space in 10 dimensions in \tIIB string theory, where for future convenience we choose cylindrical coordinates in the first three $\mathbb{R}^2$ planes and where two of the directions are periodic,
\begin{align}
  \wt x_8 &= \wt R_1\  \wt u_1 \  ,& \wt x_9 &= \wt R_2\  \wt u_2 \  .
\end{align} 
Adding the two spectator directions \( \wt x_6 \) and \( x_7 \), we have the following variables:
\begin{equation}
\rho_1, \theta_1, \rho_2, \theta_2, \rho_3, \theta_3, \wt x_6, x_7, \wt u_1, \wt u_2.
\end{equation}
In the notation of~\cite{Reffert:2011dp} we want to set up a fluxbrane with two independent deformation parameters, one of which being purely real, the other being purely imaginary.  
Shifts are induced in the $\theta_1,\ \theta_2$--directions, which for supersymmetry preservation need to be compensated by a shift in the $\theta_3$--directions. We impose the monodromies\footnote{The two parameters \( \epsilon_1 \) and \( \epsilon_2 \) are real. In the habitual conventions for the \( \Omega \)--deformation they correspond to a real and a purely imaginary \( \varepsilon \). See~\cite{Nekrasov:2010ka,Reffert:2011dp} for comparison.}
\begin{align}
   \begin{cases}
    \wt u_1 \simeq \wt u_1 + 2 \pi\ , \\
    \theta_1 \simeq \theta_1 + 2 \pi \epsilon_1 \wt R_1\ , \\
    \theta_3 \simeq \theta_3 - 2 \pi \epsilon_1 \wt R_1\ , \\
  \end{cases} &&
   \begin{cases}
    \wt u_2 \simeq \wt u_2 + 2 \pi\ , \\
    \theta_2 \simeq \theta_2 + 2 \pi \epsilon_2 \wt R_2 \ , \\
    \theta_3 \simeq \theta_3 - 2 \pi \epsilon_2 \wt R_2 \ , \\
  \end{cases}
\end{align}
and we change to new angular coordinates $\phi_i$ which are $2\pi$ periodic:
\begin{align}
  \theta_1 &=  \phi_1 + R_1\epsilon_1 \wt u_1 \ ,\\
  \theta_2 &=  \phi_2 + R_2\epsilon_2 \wt u_2 \ ,\\
  \theta_3 &=  \phi_3 - R_1\epsilon_1 \wt u_1 - R_2\epsilon_2\wt u_2 \ .
\end{align} 
This results in a fluxbrane background, where for later convenience we introduce a constant dilaton field:
\begin{equation}
  \Phi_0 = \log \frac{\alpha'}{\wt R_1 \wt R_2} \ .  
\end{equation}

\paragraph{The Fluxtrap.}
The \emph{fluxtrap} background is obtained if we T--dualize $\wt u_1$ and $\wt u_2$ into \( u_1 \) and \( u_2 \).  After a final coordinate change to eliminate \( \phi_3 \),
\begin{equation}
  \phi_1 + \phi_2 + \phi_3 = \psi,
\end{equation}
we obtain the bulk fields for the double fluxtrap background:
\begin{subequations}
  \label{eq:omega-frame}
  \begin{align}
    \begin{split}
      \di s^2 ={}& \di \rho_1^2 + \frac{\rho_1^2}{ \Delta_1^2 } \di \phi_1^2 + \di \rho_2^2 + \frac{\rho_2^2}{ \Delta_2^2 } \di \phi_2^2 + \frac{ \Delta_2^2 \di x_8^2 + \Delta_1^2 \di x_9^2 + \rho_3^2 {\left( \epsilon_2 \di x_8 - \epsilon_1 \di x_9 \right)}^2}{\Delta_1^2 \Delta_2^2 + \rho_3^2 \left( \epsilon_1^2 \Delta_2^2 + \epsilon_2^2 \Delta_1^2 \right)}  \\
      & + \di \rho_3^2 + \frac{\rho_3^2 \Delta_1^2 \Delta_2^2 }{
        \Delta_1^2 \Delta_2^2 + \rho_3^2 \left( \epsilon_1^2
          \Delta_2^2 + \epsilon_2^2 \Delta_1^2 \right)} {\left( \di
        \psi - \frac{ \di \phi_1}{\Delta_1^2 } - \frac{ \di
          \phi_2}{\Delta_2^2} \right)}^2+ \di \wt x_6^2 + \di x_7^2\ ,
    \end{split}
    \\
    \begin{split}
      B ={}& - \frac{\epsilon_1 \rho_1^2}{\Delta_1^2} \di \phi_1
      \wedge \di x_8 - \frac{\epsilon_2 \rho_2^2}{\Delta_2^2} \di
      \phi_2 \wedge \di x_9  \\
      & + \rho_3^2 \left( \di \psi - \frac{\di \phi_1}{\Delta_1^2} - \frac{\di \phi_2}{\Delta_2^2 } \right) \wedge \frac{\epsilon_1 \Delta_2^2 \di x_8 + \epsilon_2 \Delta_1^2 \di x_9 }{\Delta_1^2 \Delta_2^2 + \rho_3^2 \left( \epsilon_1^2 \Delta_2^2 + \epsilon_2^2 \Delta_1^2 \right)}\  ,
    \end{split}
\\
    \eu^{- \Phi} ={}& \sqrt{ \Delta_1^2 \Delta_2^2 + \rho_3^2 \left(
        \epsilon_1^2 \Delta_2^2 + \epsilon_2^2 \Delta_1^2 \right)}\ ,
  \end{align}
\end{subequations}
where
\begin{equation}
  \Delta_i^2 = 1 + \epsilon_i^2 \rho_i^2 \  ,  
\end{equation}
and \( x_8, x_9 \) are defined by
\begin{align}
  x_8 &= \frac{\alpha'}{\wt R_1} u_1 \ , & x_9 &= \frac{\alpha'}{\wt R_2} u_2 \  .
\end{align}
The advantage of these coordinates is that the limit
\begin{equation}
  \wt R_i \to 0  
\end{equation}
is smooth. Hence from now on we are free consider \( x_8 \) and \( x_9 \) as non-compact.  

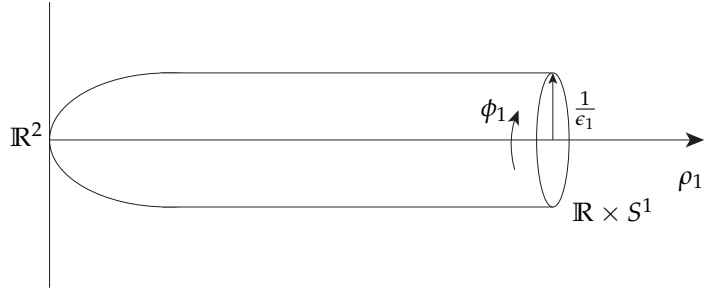
\begin{figure}
  \centering
    \begin{tikzpicture}
      \node (0,0) {\includegraphics[scale=.7,trim=0 0 0 16em, clip=true]{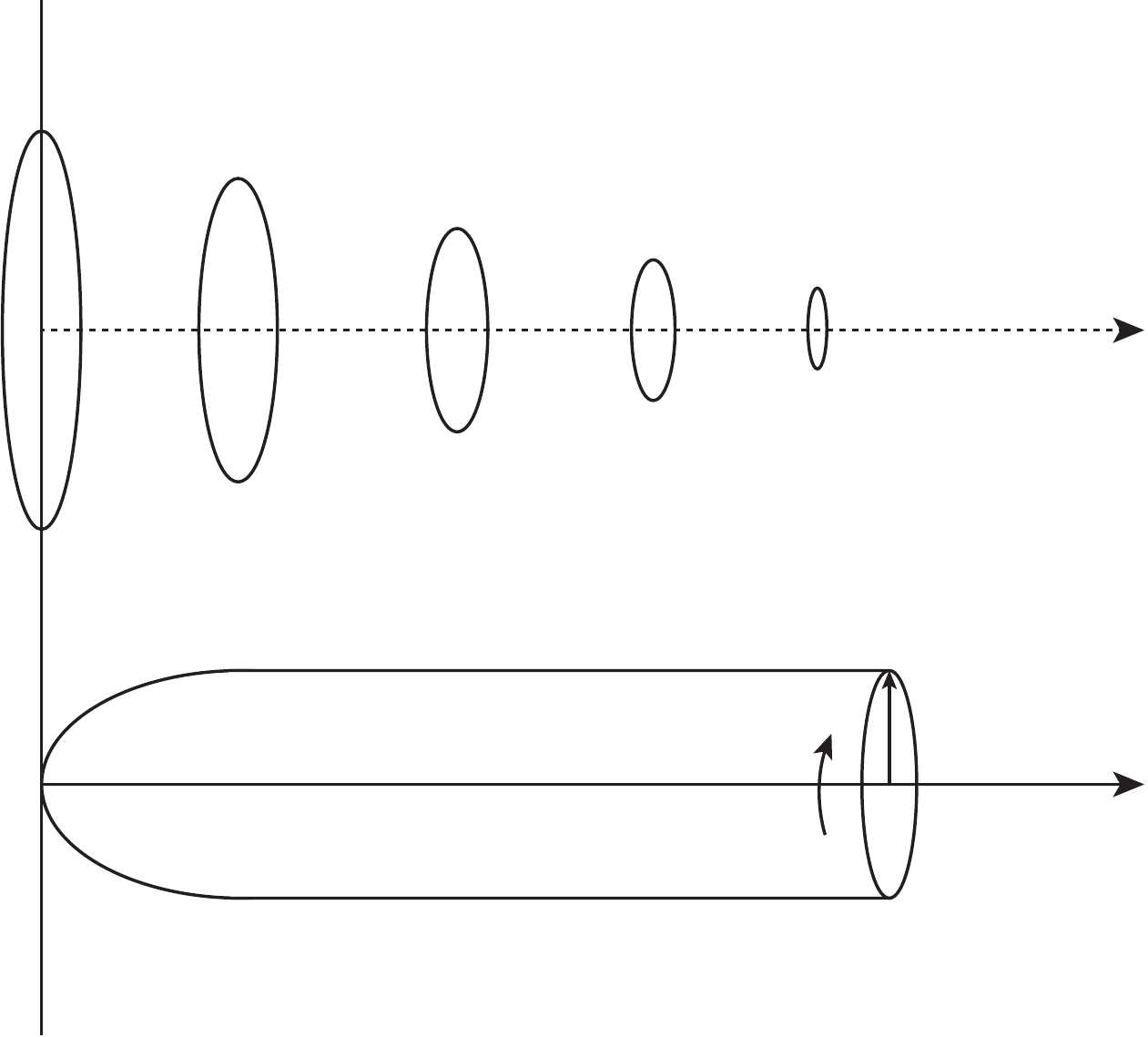}};
      \begin{scope}[shift={(-6,-2)},every node/.style={anchor=base
          west}]
        \begin{small}

          \draw (10,1.5) node[]{\( \rho_1 \)}; 
          \draw (8,2.4) node[anchor={base east}]{\( \phi_1  \)};
          \draw (8.6,2.4) node[anchor={base west}]{\( \frac{1}{\epsilon_1} \)};

          \draw (1.2,2) node[]{\( \setR^2 \)};

          \draw (8.6,1) node[]{\( \setR \times S^1 \)};

        \end{small}
      \end{scope}
    \end{tikzpicture}
  \caption{Cartoon of the geometry of the base of the manifold \( M_3( \epsilon_1) \): a cigar with asymptotic radius \( 1/ \epsilon_1 \). }
  \label{fig:cigar-fibration}
\end{figure}

In the following it will be natural to study the situation in which \( \rho_3 \muchlessthan \rho_1, \rho_2 \). In this limit, the background simplifies and it becomes easier to describe the geometry. The fields take the form
\begin{subequations}
  \label{eq:limit-fluxtrap}
  \begin{align}
    \di s^2 &= \di \rho_1^2 + \frac{\rho_1^2 \di \phi_1^2 + \di x_8^2 }{1 + \epsilon_1^2 \rho_1^2} + \di \rho_2^2 + \frac{\rho_2^2 \di \phi_2^2 + \di x_9^2 }{1+ \epsilon_2^2 \rho_2^2} + \di \rho_3^2 + \rho_3^2 \di \psi^2 + \di \wt x_6^2 + \di x_7^2 \  ,\\
    B &= \epsilon_1 \frac{\rho_1^2 }{1 + \epsilon_1^2 \rho_1^2} \di \phi_1 \wedge \di x_8 +  \epsilon_2 \frac{\rho_2^2 }{1 + \epsilon_2^2 \rho_2^2} \di \phi_2 \wedge \di x_9 \  , \\
    \eu^{- \Phi} &= \sqrt{\left( 1 + \epsilon_1^2 \rho_1^2 \right) \left(
        1 + \epsilon_2^2 \rho_2^2 \right)} \  .
  \end{align}
\end{subequations}
The space splits into a product
\begin{equation}
  M_{10} = M_3 (\epsilon_1) \times M_3 (\epsilon_2) \times \setR^3 \times S^1 \  ,  
\end{equation}
where \( \setR^3 \) is generated by \( (\rho_3, \psi, x_7) \), the \( S^1 \) is generated by \( \wt x_6 \), and \( M_3 \) is a three-dimensional manifold which is obtained as a \( \setR \) foliation (generated by \( x_8 \) or \( x_9 \)) over the cigar with asymptotic radius \( 1/\epsilon_i \) described by \( (\rho_1, \phi_1) \) or \( (\rho_2, \phi_2) \) (see the cartoon in Figure~\ref{fig:cigar-fibration}):
\begin{equation}
  \begin{tikzpicture}[node distance=5em, auto]
    \node (S) {\( \setR \langle x_8 \rangle \) }; 
    \node (M) [right of=S] {\( M_3 (\epsilon_1)\) };
    \node (cigar) [below of=M] {cigar \(\langle \rho_1, \phi_1 \rangle\)}; 
    \draw[->] (S) to node {} (M); 
    \draw[->] (M) to node {} (cigar);
  \end{tikzpicture}
\end{equation}

This shows that the effect of the \( \Omega \)--deformation is to regularize the rotations generated by \( \del_{\phi_1} \) and \( \del_{\phi_2} \) in the sense that the operators become bounded:
\begin{align}
  \| \del_{\phi_1} \|^2 &= \frac{\rho_1^2}{1 + \epsilon_1^2 \rho_1^2} < \frac{1}{\epsilon_1^2} \  ,&
  \| \del_{\phi_2} \|^2 &= \frac{\rho_2^2}{1 + \epsilon_2^2 \rho_2^2} < \frac{1}{\epsilon_2^2} \  .
\end{align}
In a different frame this will translate into a bound on the asymptotic coupling of the effective gauge theory for the motion of a D--brane.

As a final remark we observe that even though the background in Equation~(\ref{eq:limit-fluxtrap})  where the contributions of the two \( \epsilon_i \) are decoupled was obtained as a limit, it is by itself a solution of the ten-dimensional supergravity equations of motion for any value of \( \rho_i \). 

\bigskip

What we have obtained is the starting point of the chain of dualities leading eventually to the reciprocal background, as detailed in Table~\ref{tab:duality-chain}.

\paragraph{M--theory.}
As a first step we dualize in \( \wt x_6 \) to \tIIA and then lift to M--theory. A remarkable feature of the M--theory background is the fact that it is symmetric under the exchange \( \{\rho_1, \phi_1, x_8, \epsilon_1 \} \leftrightarrow \{\rho_2, \phi_2, x_9, \epsilon_2 \}\). This is the origin of the S--duality covariance of the final \tIIB background. This has to be contrasted with the fact that the directions \( x_6  \) and the M--circle \( x_{10} \) appear in a non-symmetric fashion. This is the reason why the \( \Omega \)--deformed four dimensional \( \mathcal{N} = 4 \) theory has a complicated behavior under S--duality. The explicit expression for metric and \( A_3 \) potential in the \( \rho_3 \muchlessthan \rho_1, \rho_2 \) limit is
\begin{subequations}
  \label{eq:M-frame}
  \begin{align}
    \begin{split}
      \di s^2 ={}& \left( \Delta_1 \Delta_2 \right)^{2/3} \Big[ \di \rho_1^2 + \frac{\epsilon_1^2 \rho_1^2}{1 + \epsilon_1^2 \rho_1^2} \di \sigma_1^2 + \frac{\di x_8^2 }{1 + \epsilon_1^2 \rho_1^2} + \di \rho_2^2 + \frac{\epsilon_2^2 \rho_2^2}{1 + \epsilon_2^2 \rho_2^2} \di \sigma_2^2 + \frac{\di x_9^2 }{1 + \epsilon_2^2 \rho_2^2} \\ 
      &+ \di \rho_3^2 + \rho_3^2 \di  \psi^2 + \di x_6^2 + \di x_7^2 \Big] + \left( \Delta_1 \Delta_2  \right)^{-4/3} \di x_{10}^2 \  ,
    \end{split} \\
    A_3 ={}& \frac{ \epsilon_1^2 \rho_1^2}{1+ \epsilon_1^2 \rho_1^2} \di \sigma_1 \wedge \di x_8 \wedge \di x_{10} + \frac{ \epsilon_2^2 \rho_2^2}{1 + \epsilon_2^2 \rho_2^2} \di \sigma_2 \wedge \di x_9 \wedge \di x_{10} \  ,
  \end{align}
\end{subequations}
where \( x_{10} \) is periodic with period \( 2 \pi R_{10} \) and \( R_{10} \) is related to the string coupling and string length in the fluxtrap as follows:
\begin{equation}
  R_{10} =  g_\Omega \ell_\Omega \  .
\end{equation}
We have also introduced two periodic coordinates \( \sigma_i \) using the asymptotic radii in the cigars:
\begin{equation}
  \sigma_i = \frac{\phi_i}{\epsilon_i} \  .    
\end{equation}
These are the directions in which we will reduce the \M5--brane to get to the effective description in terms of a \D3--brane in the reciprocal frame.
A final remark is needed concerning the symmetries of the background that will be reflected in the properties of the gauge theories. The coefficients of the terms \( \di \sigma_1^2 \) and \( \di \sigma_2^2 \) are interchanged under the exchange \( \epsilon_1 \leftrightarrow \epsilon_2 \), while there is no obvious symmetry between the coefficients of \( \di x_6^2 \) and \( \di x_{10}^2 \). We will see that this leads to S--duality covariance of the reciprocal theory, which is absent in the \( \Omega \)--deformed \textsc{sym}. 

\paragraph{Type IIA.}
The second step of the 9/11 flip is obtained by reducing the M--theory description to \tIIA on the coordinate \( \sigma_1 \). It is well-known that the reduction of flat space on this angle gives rise to the near-horizon limit of a \D6--brane in \tIIA~\cite{Townsend:1995af}\footnote{Flat space can be seen as the \( r \to 0 \) limit of a Taub--\textsc{nut} space.}. The same applies to our background that can be described the as the backreaction in the near-horizon limit of a \D6--brane in the fluxtrap\footnote{It was found in~\cite{Hellerman:2012zf} that the effect of the fluxtrap on the \D6 can be understood in terms of a non-commutative deformation. Non-commutativity in the \( \Omega \)--background is a topic of interest in the recent literature~\cite{Fucito:2011pn,Fucito:2012xc}. It would be interesting to relate these observations to the topic of non-commutativity in closed strings~\cite{Andriot:2012an,Lust:2012fp}.}. Note that this reduction is different from the one on the dual Melvin circle, that leads to a different realization of the \( \Omega \)--deformation discussed in~\cite{Billo:2006jm}.

The definition of the string coupling \( \gAr \) and of the string length \( \lr \) follow from the radius of the coordinate \( \sigma_1 \) that we used for the compactification. Imposing that the tension of a \D0--brane coincides with the inverse radius,  we find 
\begin{equation}
  T_{\D0} = \frac{1}{\gAr \lr} = \epsilon_1 \Rightarrow \gAr = \frac{1}{\epsilon_1 \lr} \  .  
\end{equation}
Imposing that the tension of the \M5--brane wrapped on \( \sigma_1 \) coincides with the tension of the \D4--brane in this frame, we derive the value of the Planck length \( \ell_p \):
\begin{equation}
  T_{\M5} \frac{2\pi}{\epsilon_1} = \frac{1}{32 \pi^5 \ell_p^6} \frac{2 \pi}{\epsilon_1} = T_{\D4} = \frac{1}{32 \pi^5 \gAr  \lr^5 } \Rightarrow \ell_p = \frac{\lr^{2/3}}{\epsilon_1^{1/3}} \  .
\end{equation}
The same %
condition can be imposed to find the relationship between the string length \( \ell_\Omega \) and gauge coupling in the \tIIA fluxtrap \( \gAO \). In a compactification on \( x_{10} \):
\begin{align}
  \label{eq:gIIAOmega-R10}
  T_{\D0} &= \frac{1}{\gAO \lO} = \frac{1}{R_{10}} \Rightarrow \gAr = \frac{R_{10}}{\lO} \  ,\\
\intertext{and}
  T_{\M5} 2 \pi R_{10} &= \frac{1}{32 \pi^5 \ell_p^6} 2 \pi R_{10} = T_{\D4} = \frac{1}{32 \pi^5 \gAO  \lO^5 } \Rightarrow \ell_p = \gAO^{1/3} \lO.
\end{align}
Comparing the two values for the Planck length we find
\begin{equation}
  \label{eq:length-mapping}
  \frac{\lr^2}{\epsilon_1} = \gAO \lO^3\,.
\end{equation}

\paragraph{Type IIB.}
The last step consists in a T--duality in \( \sigma_2 \). Since the T--dual of flat space in \( \sigma_2 \) is the near-horizon limit of an \NS5--brane~\cite{Gregory:1997te}, we can describe the final \tIIB background as an \( \Omega \)--deformed \NS5--\D5 system. The configuration preserves eight Killing spinors, as derived explicitly in Appendix~\ref{sec:supersymmetry}. The string coupling constant \( \gBr \) is derived from the coupling in \tIIA and the compactification radius \( 1/\epsilon_2 \):
\begin{equation}
  \boxed{\gBr = \gAr \lr \epsilon_2 = \frac{\epsilon_2}{\epsilon_1} \  . } 
\end{equation}
Once more for simplicity we report the explicit expressions of the bulk fields in this last frame (the reciprocal frame) in the 
limit where $\rho_3 \muchlessthan \rho_1,\ \rho_2$:
\begin{subequations}
  \label{eq:reciprocal-frame}
  \begin{align}
    \begin{split}
      \di s^2 ={} & \epsilon_1 \rho_1 \sqrt{1+\epsilon_2^2\rho_2^2}\left[\di \rho_1^2+\di \rho_2^2+\frac{\di \tilde \sigma_2^2} {\epsilon_1^2 \rho_1^2 \epsilon_2^2 \rho_2^2} + \di \rho_3^2+\rho_3^2\di \psi^2+\di x_6^2+\di x_7^2 +\right.\\
      & \left.+\frac{\di x_8^2}{1+\epsilon_1^2\rho_1^2}+\frac{\di x_9^2}{1+\epsilon_2^2\rho_2^2} +\frac{ \di x_{10}^2}{ \left( 1 +\epsilon_1^2\rho_1^2\right) \left( 1 + \epsilon_2^2\rho_2^2\right)}
      \right] \ ,
    \end{split}
    \\
    B ={} & \frac{ \epsilon_1^2 \rho_1^2}{1+\epsilon_1^2\rho_1^2} \di x_8 \wedge\di x_{10} \ , \\
    \eu^{- \Phi} ={} & \frac{\epsilon_2 \rho_2}{\epsilon_1 \rho_1} \sqrt{\frac{1+ \epsilon_1^2 \rho_1^2}{1 + \epsilon_2^2 \rho_2^2}} \ , \\
    C_2 ={} & \frac{\epsilon_2^2 \rho_2^2}{ 1 + \epsilon_2^2 \rho_2^2} \di x_9 \wedge\di x_{10} \  ,
\end{align}
\end{subequations}
where \( \tilde \sigma_2 \) is periodic with period \( 2 \pi \alpha' \epsilon_2 \).  The fluxes that appear here are due to the fluxtrap construction and are not  the ones generated by the background branes, which are negligible in the \( \rho_3 \muchlessthan \rho_1, \rho_2 \) limit that we are considering. An interesting feature of this background is that the dilaton vanishes asymptotically for \( \rho_1, \rho_2 \to \infty \). We will use this fact in the study of the gauge theory to identify the value of the effective gauge coupling.
 
As anticipated from the M--theory %
description, the \tIIB background has a simple behavior under S--duality which %
amounts
to exchanging \( \epsilon_1 \) with \( \epsilon_2 \).  This transformation has the effect of swapping the \NS5--brane with the \D5--brane in the bulk.

\section{The weakly coupled theories on the D3--brane}
\label{sec:weekly-coupl-theory}

Now that the fluxtrap background is set up, we want to consider the full configuration, including the branes that will lead us to an effective theory in the two \tIIB duality frames. 

Conceptually we are starting from the M--theory picture with an \M5--brane extended in \( (\rho_1, \rho_2, \sigma_1, \sigma_2, x_6, x_{10}) \), \emph{i.e.} on \( \setR_+^2 \times T^2 \times T^2 \) in the bulk of Equation~\eqref{eq:M-frame}. The complex structures of the two tori are respectively \( \widehat \tau = \im \epsilon_1 / \epsilon_2 \) and \( \tau = \im / \gBO \). The compactification on the first torus gives the \( \Omega \)--deformed \( \mathcal{N} = 4 \) super-Yang--Mills with coupling \( \gYM^2 = 2 \pi \im / \tau \) and deformation parameters \( \epsilon_1  \) and \( \epsilon_2 \); the compactification on the second torus leads to the reciprocal theory with coupling \( \grec^2 = 2 \pi \im / \widehat \tau = 2 \pi  \epsilon_2 / \epsilon_1 \) on a torus foliation \( T^2 \to \setR^2 \) with complex structure \( \tau = 2 \pi \im / \gYM^2 \).
Both gauge theories can be obtained as limits of the deformed \( (2,0) \) six-dimensional gauge theory and in particular inherit four conserved supercharges (see Appendix~\ref{sec:supersymmetry}). In practice it is computationally more convenient to describe the effective actions for the \D3--branes in the fluxtrap of Equation~\eqref{eq:omega-frame} and in the reciprocal frame of Equation~\eqref{eq:reciprocal-frame}.

\paragraph{The \( \Omega \)--deformed \( \mathcal{N}=4 \) \textsc{SYM}.}
The effective theory for a Hanany--Witten setup of \D3--branes suspended between \NS5--branes in the fluxtrap background reproduces the \( \Omega \)--deformation of \( \mathcal{N} = 2 \) \ac{sym}~\cite{Hellerman:2012zf}. Here we wish to describe the deformation of \( \mathcal{N} = 4 \) gauge theory, which is the effective description of a stack of \( N \)  parallel \D3--branes extended in \( (\rho_1, \phi_1, \rho_2, \phi_2 )\) in the fluxtrap background\footnote{The \( \Omega \)--deformation of \( \mathcal{N} = 4 \) \ac{sym} is different from the one proposed in~\cite{Ito:2012hs}. In our language, the latter results from identifications and T--dualities in all the six directions orthogonal to the \D3--brane.}. A major difference with the \( \mathcal{N} = 2 \) case is that now the \D3--brane can move in six directions, which is conveniently expressed in three complex fields \( (\varphi, z, w) \). 
Consider the static embedding of a \D3--brane extended in \( (\rho_1, \phi_1, \rho_2, \phi_2 )\) and described by flat coordinates \( \xi^i \), moving in the directions
\begin{align}
  w(\xi) &= \frac{\rho_3 \eu^{\im \psi}}{\pi \alpha'}\ , & z(\xi) &= \frac{x_6 + \im x_7}{ \pi \alpha'}\ , & \varphi(\xi) &= \frac{x_8 + \im x_9}{ \pi \alpha'} \ .
\end{align}
For \( N = 1 \), the dynamics is given by the \ac{dbi} action:
\begin{multline}
  \mathscr{L}_{\Omega} = \frac{1}{4\gYM^2}\Bigg[ F_{ij}F^{ij} + \frac{1}{2} \left( \del^i \varphi + V^k F\indices{_{k}^{i}}\right) \left( \del_i \bar \varphi + \bar V^kF_{ki} \right)  - \frac{1}{8}{( \bar V^i \del_i \varphi - V^i \del_i \bar \varphi + V^k\bar V^l F_{kl} )}^2 \\
 + \frac{1}{4} \left( \delta^{ij} + V^i \bar V^j \right) \left( \del_i z \del_j \bar z + \text{c.c.} \right) + \frac{1}{4} \left( \delta^{ij} + V^i \bar V^j \right) \left( \del_i w \del_j \bar w + \text{c.c.} \right)  \\
 + \frac{1}{2 \im}  \left( \epsilon_3 \bar V^i + \bar \epsilon_3 V^i \right)  \left( \bar w \del_i  w - \text{c.c.} \right) + \frac{1}{2} \abs{\epsilon_3}^2 w \bar w \Bigg] \  ,
\end{multline}
where \( V  = \epsilon_1 \left( \xi^0 \del_1 - \xi^1 \del_0 \right) + \im \epsilon_2 \left( \xi^2 \del_3 - \xi^3 \del_2 \right) \) and \( \gYM^2 = 2 \pi \gBO \). 
The action is expanded up to second order in the derivatives and we find that the highest term in $\epsilon$ is of order $\mathcal{O}(\epsilon^4)$.
In comparing with the \( \mathcal{N} = 2 \) case~\cite{Nekrasov:2003rj,Ito:2010vx}, we see that the new scalar fields $z$ and $w$ have an unusual kinetic term $ \left( \delta^{ij} + V^i \bar V^j \right)$, moreover $w$ has a term with one derivative due to the breaking of Lorentz invariance, and a mass proportional to $\abs{\epsilon_3}=\sqrt{\abs{\epsilon_1}^2+\abs{\epsilon_2}^2}$.

\def\LOCVAR{U}
\paragraph{The weakly coupled reciprocal theory.}
Following the chain of dualities, the \D3--brane turns first into a \D4--brane in \tIIA, and then into an \M5--brane extended in \( (\rho_1, \sigma_1, \rho_2, \sigma_2, x_6, x_{10}) \) in M--theory. The reduction to \tIIA turns the \M5 into a \D4--brane and the T--duality finally leads to a \D3--brane in the reciprocal background (see Table~\ref{tab:Everything}). The effective theory of this brane is what we call the reciprocal gauge theory.
Consider the static embedding for the D--brane extended in \( \rho_1, \rho_2, x_6, x_{10} \):%
\begin{align}
  \rho_1 &= y_1 \ , & \rho_2 &= y_2 \ , & x_6 &= y_3 \ , & x_{10} &= y_4 \ .
\end{align}
The geometry seen by the \D3--brane is of a two-torus fibration (generated by \( y_3, y_4 \)) over \( \setR^2_+ \)  (generated by \( y_1, y_2 \))
\begin{equation}
  \begin{tikzpicture}[node distance=5em, auto]
    \node (T2) {\( T^2 \langle y_3, y_4 \rangle\) }; 
    \node (M) [right of=S] {\( M_4 \) };
    \node (R2) [below of=M] { \( \setR^2_+ \langle y_1, y_2 \rangle \)}; 
    \draw[->] (T2) to node {} (M); 
    \draw[->] (M) to node {} (R2);
  \end{tikzpicture}
\end{equation}

The dynamics is described by the fields
\begin{align}
  \LOCVAR_1 + \im \LOCVAR_2 &= \frac{\rho_3 \eu^{\im \psi}}{2 \pi \alpha'} \ , & \LOCVAR_3 &= \frac{x_7}{2 \pi \alpha'} \ , & \LOCVAR_4 &= \frac{\tilde \sigma_2 }{2 \pi \alpha'} \ , & \LOCVAR_5 &= \frac{x_8}{2 \pi \alpha'} \ , & \LOCVAR_6 &= \frac{x_9}{2 \pi \alpha'} \  .
\end{align}
The effective action for the \D3--brane is
\begin{multline}
  \mathscr{L}_{\text{rec}} = \frac{y_2}{8 \pi y_1} F_{kl} F_{kl} + \frac{\epsilon_1^2 y_1 y_2}{4 \pi} \Bigg[ \sum_{k=1}^3 {\left( F_{k4} - \del_k \LOCVAR_5 \right)}^2 + \frac{1}{\Delta_2^2} \sum_{k=1}^3 {\left( \im \frac{\epsilon_2}{\epsilon_1} \frac{y_2}{y_1} {(*F)}_{k4} - \del_k \LOCVAR_6 \right)}^2 \\ 
+ \tau^{kl} (\xi) h^{ij} (\xi) \del_k \LOCVAR_i \del_l \LOCVAR_j  +  \Delta_2^2 {(\del_4 \LOCVAR_5)}^2 + \Delta_1^2 {( \del_4 \LOCVAR_6)}^2 + \left( y_1^{-2} + y_2^{-2} \right) \left( \LOCVAR_1^2 + \LOCVAR_2^2 \right) \Bigg],
\end{multline}
where
\begin{align}
  \tau^{kl}(\xi) &=
  \begin{pmatrix}
    1 \\ & 1 \\ && 1 \\ &&& \Delta_1^2 \Delta_2^2
  \end{pmatrix} \  ,&
 h^{ij}(\xi) &=
  \begin{pmatrix}
    1 \\ & 1 \\ && 1 \\ &&& {\left(\epsilon_1 y_1\right)}^{-2} {\left(\epsilon_2 y_2\right)}^{-2}
  \end{pmatrix}.
\end{align}

The couplings to $F$ are inherited from the $B$--field, while the couplings to the dual $(*F)$ are inherited from $C_2$ via the Chern--Simons term. The two are interchanged under S--duality as we will see in the following. The dilaton appears in the effective gauge coupling that will be evaluated below. Lorentz invariance is broken as a result of the asymmetry in the directions $x_6$ and $x_{10}$. 

In the previous section, we have seen that the bulk is the backreaction of the near-horizon limit of an NS5-- and a D5--brane. This introduces an issue concerning the boundary conditions of the gauge theory at $\rho_1=0$ and $\rho_2=0$. In the low-energy limit of the gauge theory under consideration, the boundary conditions can be understood as local defect operators at the origin. As we will see in the following, the generic \ac{bps} state in this frame lives away from the origin, so that the issue of these boundary conditions is not central for our construction. Nevertheless, this deserves future detailed analysis in connection with the states contributing to the full partition function for the \(\Omega\)--deformed theory.%
\footnote{A configuration very similar to ours was described in~\cite{Yagi:2012xa}, where the author argues that it is possible to recover the dynamics of a chiral gauged \textsc{wzw} model from the boundary couplings. Given the known difficulties in defining the ``chiral half of Liouville theory'' stemming from anomalies and the presence of states with fractional spin, we believe that the proposed explicit string realization of the \ac{agt} correspondence deserves further analysis.} 

The reciprocal four-dimensional theory has broken Lorentz and translational invariance. This means that we need to address issues related to that, which do not arise in Lorentz-invariant backgounds.  In particular we would like to examine the gauge coupling and its behavior under S--duality, but to do this, we must define what we mean by ``the gauge coupling'' in a background with so much broken symmetry. For this purpose it is convenient to set all the scalars to zero and concentrate on the gauge part of the action:
\begin{equation}
  \mathscr{L}_{g}(\epsilon_1, \epsilon_2) = \frac{y_2}{4 \pi y_1} \left[ \left( 1 + \epsilon_1^2 y_1^2 \right) F_{4k} F_{k4} + \frac{(*F)_{k4} {(*F)}_{k4}}{ 1 + \epsilon_2^2 y_2^2} \right]  \  .
\end{equation}
There is no unique definition of the gauge coupling for an action in which Lorentz invariance is broken and the gauge kinetic term is not diagonal in \( F_{ij} \). It is convenient to define the gauge kinetic \emph{tensor} \( M^{ijkl} \) from
\begin{equation}
  \mathscr{L}_g =  M^{ijkl} F_{ij} F_{kl} \  ,
\end{equation}
where \( M \in K = \Lwe^2 \setR^4 \otimes \Lwe^2 \setR^4 \). 

\paragraph{Gauge coupling.}

If we introduce the natural inner product on \( \Lwe^2 \setR^4 \),
\begin{equation}
  \begin{aligned}
    \scal{\cdot}{\cdot}_{\Lwe} : \Lwe^2 \setR^4 \times \Lwe^2 \setR^4 &\to \setR \\
    (a, b) &\mapsto \scal{a}{b}_{\Lwe} = a^{ij} b^{kl} \epsilon_{ijkl}\,,
  \end{aligned}
\end{equation}
then \( K \) inherits the inner product as
\begin{equation}
  \begin{aligned}
    \scal{\cdot}{\cdot}_K : K \times K &\to \setR \\
    ( a_1 \otimes a_2 , b_1 \otimes b_2 ) &\mapsto \scal{a_1 \otimes a_2}{b_1 \otimes b_2 }_K = \scal{a_1}{b_1}_{\Lwe} \scal{a_2}{b_2}_{\Lwe} = a_1^{ij} b_1^{kl} \epsilon_{ijkl} a_2^{i'j'} b_2^{k'l'} \epsilon_{i'j'k'l'}\,.
  \end{aligned}
\end{equation}
We can now define the scalar \( \geff \) in terms of the norm of the gauge kinetic tensor:
\begin{equation}
  \frac{1}{\geff^2} = \sqrt{ \frac{2}{3} } \norm{M}_K   = \sqrt{ \frac{2}{3} \epsilon_{ijkl} \epsilon_{i'j'k'l'} M^{iji'j'} M^{klk'l'} } \  ,
\end{equation}
where the normalization has been chosen such that \( \geff = \gYM \) in the standard Lorentz-invariant case
\begin{equation}
  \mathscr{L} = \frac{1}{4 \gYM^2} F_{ij}F_{ij} \  .
\end{equation}
In our case, we find that the effective gauge coupling of the reciprocal theory is
\begin{equation}
  \frac{1}{\grec^2} = \frac{1}{2 \pi} \frac{y_2 \sqrt{1 + \epsilon_1^2 y_1^2}}{y_1 \sqrt{1 + \epsilon_2^2 y_2^2} }.
\end{equation}
We recognize the dilaton of Equation~\eqref{eq:reciprocal-frame} in the reciprocal frame. In the large--\( y \) limit, \emph{i.e.} far away from the singularity, this reduces to
\begin{equation}
  \boxed{\grec^2 \xrightarrow[y_1, y_2 \to \infty]{} 2 \pi \frac{\epsilon_2}{ \epsilon_1} \  . }
\end{equation}
We see that the asymptotic gauge coupling is given by the ratio of the two $\epsilon$--parameters as in the Liouville theory in the \ac{agt} correspondence.

\paragraph{S--duality.}

In order to study the behavior of the action under S--duality we need a notion of inverse coupling. For this purpose we can look at \( K \) as the set of linear operators \( \Omega^2(\setR^4) \to \Lwe^2 \setR^4 \), \emph{i.e.} as the set of square matrices acting on the vector space \( \setR^6 \), and define \( M^{-1} \in K \) as the inverse matrix to \( M \). Then we can define the S--dual to the action
\begin{equation}
  \mathscr{L}_g =  M^{ijkl} F_{ij} F_{kl} 
\end{equation}
as the action obtained by inverting the %
tensor \( M \) and dualizing the gauge field:
\begin{equation}
  \mathscr{L}_{\text{dual}} = \frac{1}{16 \pi^2} ( M^{-1} )^{ijkl} (*F)_{ij} (*F)_{kl} \  .  
\end{equation}

In our case this is particularly simple because \( M \) is a symmetric matrix and the action has been written explicitly in terms of the gauge field and its dual. It follows that
\begin{equation}
  \mathscr{L}_{\text{dual}}(\epsilon_1, \epsilon_2) = \frac{y_1}{4 \pi y_2} \left[ \frac{{(*F)}_{4k} {(*F)}_{k4}}{1 + \epsilon_1^2 y_1^2 } + F_{k4} F_{k4}\left( 1 + \epsilon_2^2 y_2^2 \right) \right]  .
\end{equation}
It is immediate to see that the effect of S--duality is simply to exchange \( \epsilon_1 \) and \( \epsilon_2 \) as we had already observed at the string level by looking at the reciprocal frame:
\begin{equation}
  \boxed{\mathscr{L}_{\text{dual}} (\epsilon_1, \epsilon_2) = \mathscr{L}_g(\epsilon_2, \epsilon_1) \  .  }
\end{equation}

In the \ac{agt} correspondence one identifies the Liouville parameter \( b \) with the ratio of the two epsilons,
\begin{equation}
  b^2 = \frac{\epsilon_2}{\epsilon_1} \ .  
\end{equation}
Even though the reciprocal gauge theory is intrinsically four-dimensional, we have thus seen that it shares at least two remarkable properties with the two-dimensional Liouville field theory:
\begin{enumerate}
\item The asymptotic coupling constant is proportional to \( b^2 \);
\item S--duality exchanges \( b \leftrightarrow 1/b \),  just like the Liouville duality that exchanges the perturbative and the instanton spectrum.
\end{enumerate}
Observe that these properties do not depend on the number of dynamical \D3--branes. It follows that the \( b \leftrightarrow 1/b \) symmetry exists also for the more general cases of Toda field theories, in perfect agreement with the results of the two-dimensional analysis~\cite{Kausch:1992ja}.

It is thus clear that the above construction is a first important step towards the complete recreation of the \ac{agt} correspondence within string and M--theory.

\section{BPS states and DOZZ factors}
\label{sec:BPSandDOZZ}

\def\AGT{Alday:2009aq}
\def\NOOm{Nekrasov:2003rj}
\def\SHRefflandoA{Hellerman:2011mv}
\def\SHRefflandoB{Hellerman:2012zf}
\def\JoeBook{Polchinski:1998rr}

\def\osc{oscilloid\xspace}
\def\Osc{Oscilloid\xspace}
\def\dozz{\textsc{dozz}oid\xspace}
\def\oscs{oscilloids\xspace}
\def\Oscs{Oscilloids\xspace}
\def\dozzs{\textsc{dozz}oids\xspace}

There are two sets of \ac{bps} objects that are of central importance in generating the
quantum effective action of the $\O$--deformed gauge theory\footnote{More general \ac{bps} states are possible for special values of \( \epsilon \). See for example~\cite{Bulycheva:2012ct} for an exhaustive study of the \ac{ns} limit.}.  The first are the
\ac{bps} instanton configurations, that are localized at the origin.  The second are the 
perturbative modes of the fundamental fields carrying angular
momentum along the $U(1) \times U(1)$ rotational isometries.  Mapping these contributions
to the reciprocal theory, we find that the roles of the perturbative configurations and
nonperturbative states are reversed.   As anticipated by \ac{agt}, the partition function over
instantons in the Nekrasov--Okounkov gauge theory
maps to the free-field determinant of the massless gauge field degrees of
freedom in the reciprocal theory, giving rise to the usual modular form defining the holomorphic factor of the Liouville--Toda partition function.  The perturbative \ac{bps} modes of the massive vector
multiplet, on the other hand, map in the reciprocal theory to nonperturbative states bound
to electric fluxes,
the resummation of whose virtual effects 
reproduces the holomorphic 
\textsc{dozz} factors of the Liouville--Toda partition function.  We comment briefly
on the ``reciprocal'' relationship between the perturbative and
nonperturbative \ac{bps} states in the two frames. 

In the following subsections, we will trace the string-theoretic description of the
\ac{bps} states through each
duality frame from the Nekrasov--Okounkov $\O$--deformed gauge theory with coupling 
constant $\gYM\sqd$ and deformation parameters $\e\ll i$ 
to the reciprocal gauge theory with gauge coupling $2 \pi \e\ll 2 / \e\ll 1$ and complex structure
$\t = 2 \pi \im / \gYM^2$ for the toroidally compactified directions.

There are two classes of important \ac{bps} states: the instanton-particles (instantons in four dimensions,
particles when lifted on the $x_ 6$ time-circle to five dimensions) and the angular momentum modes
of the massive vector multiplet.  As we shall be dualizing these objects several times,
it is desirable to give them more duality-invariant designations, so we will refer to them
as the \oscs and the \dozzs, respectively.

\subsection{String theory of the BPS states in the $\O$--deformed gauge theory}

In the original duality frame the masses of the \ac{bps} states (with the direction $x_ 6$
taken to be the timelike direction) can be computed either from string theory or
directly from the action.  The latter is simpler and isolates the relevant degrees of
freedom of the decoupled theory, but the latter makes the subsequent duality  
transformations more clear.  We shall do both.  The field theoretic section contains
no new content, and simply rehearses the insights of~\cite{\AGT}; however we do this
in order to give a uniform presentation with the string-theoretic description of
the same states in the original and successively dualized frames.

\colorlet{colorbrane}{white}
\colorlet{colorBPS}{gray!20}
\colorlet{colorbulk}{gray!40}

\begin{table}
  \centering
  \begin{tabular}{llccccccccccc}
    \toprule
    frame                                                                   & object   & $\rho_1$  & $\phi_1$       & $\rho_2$  & $  \phi_2$         & $\rho_3$ & $\psi$ & $x_6$      & $x_7$     & $x_8$     & $x_9$     & $x_{10}$       \\  \midrule
\rowcolor{colorbrane}                                                       & \D4      & $\times$  & $\times$       & $\times$  & $\times$           &          &        & $\times$   &           &           &           & $\blacksquare$ \\
\rowcolor{colorBPS} \cellcolor{white}                                       & F1       &           &                &           & $\circlearrowleft$ &          &        & $\times$   &           &           & $\times$  & $\blacksquare$ \\ 
\rowcolor{colorBPS} \cellcolor{white} \multirow{-3}*{\textsc{iia} fluxtrap} & \D0      &           &                &           &                    &          &        & $\times$   &           &           &           & $\blacksquare$ \\ 
\midrule
\rowcolor{colorbrane} \cellcolor{white}                                     & \M5      & $\times$  & $\times$       & $\times$  & $\times$           &          &        & $\times$   &           &           &           & $\times$       \\
\rowcolor{colorBPS} \cellcolor{white}                                       & \M2      &           &                &           & $\circlearrowleft$ &          &        & $\times$   &           &           & $\times$  & $\times$       \\
\rowcolor{colorBPS} \cellcolor{white}\multirow{-3}*{M--theory}              & momentum &           &                &           &                    &          &        & $\nearrow$ &           &           &           & $\nearrow$     \\ \midrule
\rowcolor{colorbrane} \cellcolor{white}                                     & \D3      & $\times $ & $\blacksquare$ & $\times $ &                    &          &        & $\times $  &           &           &           & $\times$       \\ 
\rowcolor{colorBPS} \cellcolor{white}                                       & \D3      &           & $\blacksquare$ &           & $\times $          &          &        & $\times$   &           &           & $\times$  & $\times$       \\
\rowcolor{colorBPS} \cellcolor{white}                                       & momentum &           & $\blacksquare$ &           &                    &          &        & $\nearrow$ &           &           &           & $\nearrow$     \\ 
\rowcolor{colorbulk} \cellcolor{white}                                      & \D5      &           & $\blacksquare$ & $\times $ &                    &          &        & $\times $  & $\times $ & $\times $ & $\times $ & $\times$       \\ 
\rowcolor{colorbulk} \cellcolor{white}\multirow{-5}*{reciprocal frame}      & \NS5     & $\times $ & $\blacksquare$ &           &                    &          &        & $\times $  & $\times $ & $\times $ & $\times $ & $\times$       \\ 
    \bottomrule
  \end{tabular}
  \caption{Extended objects in the various different frames. The objects are extended in the direction of the crosses (\( \times \)). Angular momentum in a direction is marked as $\circlearrowleft$ and momentum as $\nearrow$. The direction marked with a square (\( \blacksquare \)) in type~\textsc{ii} is not geometrical. The white background is for the dynamical branes described by the gauge theories; the light grey ({\color{colorBPS} $\blacksquare$}) for the branes that correspond to the \ac{bps} excitations and the dark grey background ({\color{colorbulk} $\blacksquare$}) for non-dynamical objects in the bulk which only appear as a consequence of the reduction from M--theory and duality along angular directions.}
  \label{tab:Everything}
\end{table}

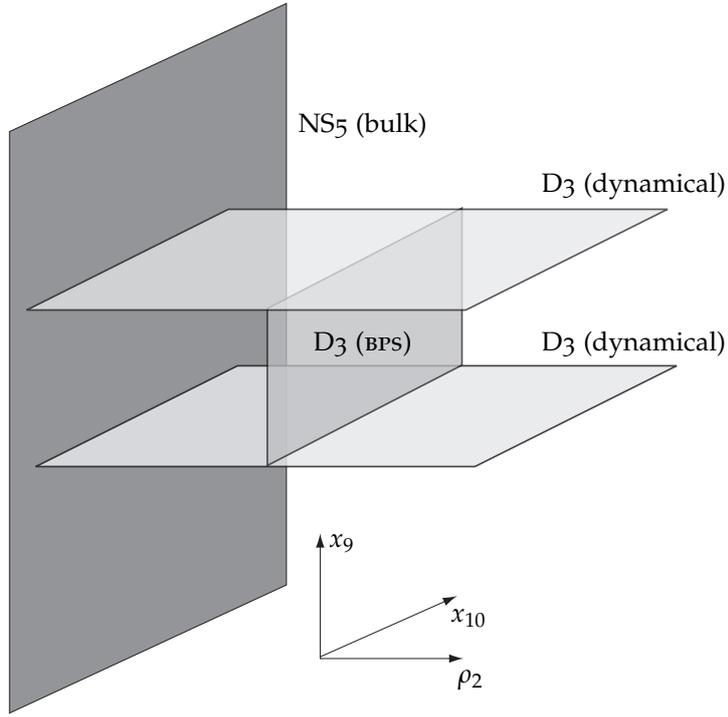
\begin{figure}
  \centering
    \begin{tikzpicture}
      \node (0,0) {\includegraphics[width=.6\textwidth]{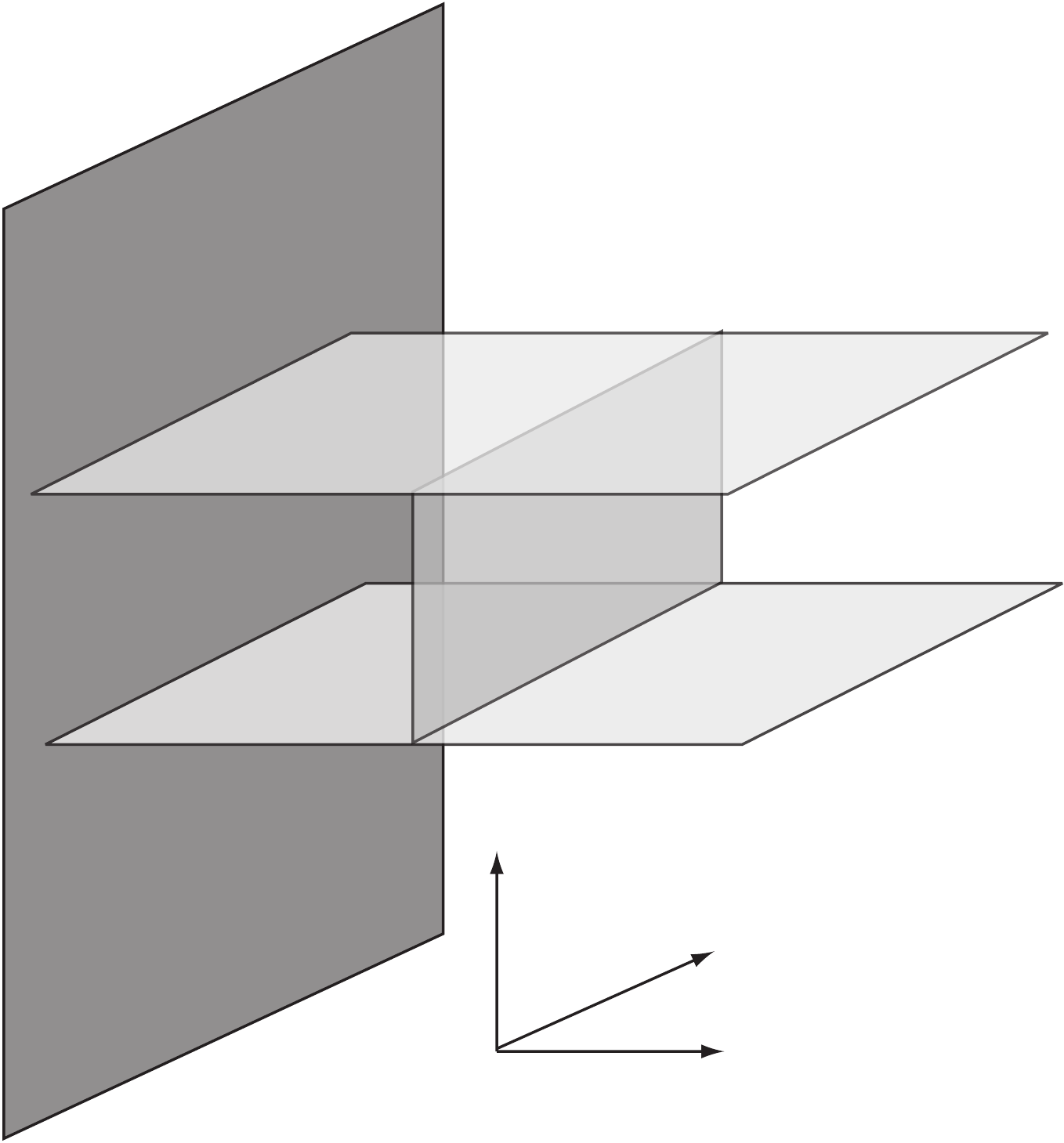}};
      \begin{scope}[every node/.style={anchor=base west}]
        \begin{small}
          \draw (-.3,-2.5) node[]{\( x_9 \)};
          \draw (1.3,-3.5) node[]{\( x_{10} \)};
          \draw (1.4,-4.3) node[]{\( \rho_2 \)};

          \draw (2.5,2.2) node[]{\D3 (dynamical)};
          \draw (2.5,0.1) node[]{\D3 (dynamical)};

          \draw (-.5,0.1) node[]{\D3 (\ac{bps})};

          \draw (-.7,3) node[]{\NS5 (bulk)};
        \end{small}
      \end{scope}
    \end{tikzpicture}
  \caption{Brane cartoon of the extended objects in the reciprocal frame. The dynamical \D3--branes end on the bulk \NS5. The \ac{bps} excitations of the gauge theory are realized as \D3--branes stretching between the two dynamical ones.}
  \label{fig:brane-cartoon}
\end{figure}

\paragraph{Field-theoretic description of the BPS \oscs.}

In the field-theoretic description in terms of four-dimensional gauge theory on 
$\O$--deformed $\IR\uu 4$ times a circle, the \ac{bps} \oscs are simply
instantons of the four-dimensional gauge theory lifted as particles in five dimensions, that are static in the (Euclidean) timelike 
$x_ 6$ direction.  We need go into this aspect no further; this description has
been discussed in detail in~\cite{\AGT, \NOOm}.  We note simply
that the localization of these objects to the origin by the $\O$--deformation is easy to
understand at the field-theoretic level: the scalar effective gauge coupling
that controls the mass of a small instanton has a global maximum at the origin, and
so the instanton's action is globally minimized there.  Further discussion of this
effect can be found in~\cite{\SHRefflandoB}.

\paragraph{Field-theoretic description of the BPS DOZZoids.}

These are linearized eigenmodes of the 4D massive vector multiplet in the 
$\O$--deformed gauge theory, with or
without angular momenta $J\ll{1,2}$ in the $U(1) \times U(1)$ angular directions.  The eigenvalue
of the Laplace operator on these modes $\chi$ obeys a \ac{bps} condition
\bbb
- \nabla\sqd \chi = \abs{\frac{L}{2\pi\apr} + \e\ll 1 J\ll 1 + \e\ll 2 J\ll 2}\sqd \cc \chi\ , \llsk\llsk
\eee
where $L$ is the distance between the branes in string frame,
and so $\frac{L}{2\pi\apr}$ is the vev of the scalar in the vector multiplet.  The angular
momenta $J\ll{1,2}$ include both orbital contributions depending on the profile of the
mode and intrinsic contributions depending on the representation of the field.
Treated
as particles in the five-dimensional gauge theory on a circle, these are \ac{bps} particle excitations
of the vector multiplet, with a mass determined by the free-field equation of motion:
\bbb
m\ll{\rm BPS} = \abs{\frac{L}{2\pi\apr} + \e\ll 1 J\ll 1 + \e\ll 2 J\ll 2 }\ . \llsk 
\eee

\paragraph{String-theoretic description of the BPS \oscs.}

In the string-theoretic description, the \ac{bps} instantons of four-dimensional gauge theory
are D-(-1)-branes of \tIIB string theory bound to D3--branes in the \tIIB fluxtrap solution
\cite{\SHRefflandoA}.  Lifted to particle-like objects of five-dimensional gauge theory on a circle,
they are D0--branes static with respect to the Euclidean timelike direction $x_ 6$. The \ac{dbi} Lagrangian for such a brane in \( \rho_3 = 0 \) is given by
\begin{equation}
  \mathscr{L} = - \mu_0 \eu^{-\Phi} = - \frac{1}{\gAO \lO} \sqrt{ \left( 1 + \epsilon_1^2 \rho_1^2 \right) \left( 1 + \epsilon_2^2 \rho_2^2 \right) } \  ,
\end{equation}
where we have used the expression for the dilaton in the fluxtrap of Equation~\eqref{eq:limit-fluxtrap}. If follows that the energy for \( n_{\D0} \) branes is
\begin{equation}
  \label{eq:D0-energy}
  \eng_{\D0} = \frac{n_{\D0}}{\gAO \lO}  \sqrt{ \left( 1 + \epsilon_1^2 \rho_1^2 \right) \left( 1 + \epsilon_2^2 \rho_2^2 \right) } \  ,
\end{equation}
which is minimized for \( \rho_1 = \rho_2 = 0 \). As already observed in~\cite{Hellerman:2012zf}, we see that these particles are localized to the origin by the spatial profile of the dilaton.

\paragraph{String-theoretic description of the BPS DOZZoids.}

Let us examine the worldsheet description of the \ac{bps} open string states stretching between two D4--branes,
and carrying angular momentum in the $U(1) \times U(1)$ rotational isometry
directions.

We begin by writing the general string worldsheet action in conformal
gauge, in the conventions of \cite{\JoeBook}.  The action is
\bbb
S
 =  \int \cc d\sqd \s \cc \mathscr{L}
 \ , 
\xxn
\mathscr{L}
 \equiv {1\over{4\pi\apr}}\cc \lsqq  -
G\ll{\m\n}(X) (\pp\ll a X\uu\m)(\pp\uu a X\uu\n) +  \e\uu{ab} \cc 
B\ll{\m\n}(X) \cc \pp\ll a X\uu\m \pp\ll b X\uu\n - \apr \cc \Phi(X) \cc \mathrm{Ric}\ll 2 
\cc\rsqq
\ .
\eee
In the \tIIA fluxtrap frame, the relevant terms in the string-frame metric, B--field
and dilaton are as in \eqref{eq:limit-fluxtrap}:
\begin{subequations}
  \begin{align}
    \di s\sqd\ll{\text{ string}} &= - \di x\ll 0\sqd + \sum\ll {i=
      1}\uu 2
    {{\r\ll i\sqd\cc \di  \phi\ll i\sqd + \di y\ll i\sqd }\over{1 + \e\ll i\sqd\cc \r\ll i\sqd}} \ , \\
    B &= \sum\ll {i= 1}\uu 2\cc \e\ll i\cc \ffk {\r\ll i\sqd\cc \di
      y\ll i \ww \di \phi\ll i} { 1 + \e\ll i\sqd
      \r\ll i\sqd} \ , \\
    \Phi &= \Phi\ll 0 - \sum\ll {i= 1}\uu 2\cc \hh\cc {\ln} ( 1 +
    \e\ll i\sqd \cc\r\ll i )\ ,
  \end{align}
\end{subequations}
where we have defined 
\begin{align}
  x_0 &= - \im x_6 \ , & y_i & = x_{i+7}  \ .
\end{align}
Classically, the string must obey the equations of motion and classical
Virasoro constraints 
\bbb
0 = T\ll{\pm\pm} = \mathcal{H}\pm \mathcal{P}\ll{\s\uu 1}\ ,
\eee
with
\bbb
\mathcal{H} \equiv  \Pi\ll\m  X\uu\m\ll{,\s\uu 0}  - \mathscr{L} - {1\over{4\pi}}\cc
(\pp\ll {\s\uu 0}\sqd 
+ \pp\ll{\s\uu 1}\sqd)\cc \Phi(X)\ , 
\een{CalHClassical}
\bbb
\mathcal{P}\ll{\s\uu 1} = - \Pi\ll\m X\uu\m\ll{,\s\uu 1}
- {1\over{4\pi}}\cc (\pp\ll {\s\uu 0} \pp\ll{\s\uu 1} )\cc \Phi(X)
 \ ,
\xxn
\Pi\ll\m \equiv {{\d \mathscr{L}}\over{\d X\uu\m\ll{,\s\uu 0}}}\ .
\eee
Now let us make the following \emph{ansatz} for a classical string trajectory:
\begin{equation}
\label{Ansatz}
\begin{aligned}
  \phi\ll {1,2} &= \phi\ll{1,2} (\s\uu 0)  = \ffk { j\ll{1,2} \cc \s\uu 0} {\rws}  \ , & X\uu 0 &= X\uu 0 (\s\uu 0) = e\uu 0 \cc \ffk {\s\uu 0} {\rws}   \ , & y\ll{1,2} &=  y\ll{1,2}(\s\uu 1) = {{L\ll{1,2}}\over{\pi \rws}} \s\ll 1 \ , \\
  X\uu 7 &= \r\ll{1,2} = \text{constant} \ , & \r\ll 3 &= 0\ ,
\end{aligned}
\end{equation}
where the distance between the branes in the $y\ll{1,2}$ directions is $L\ll{1,2}$
and the extent of the $\s\ll 1$ coordinate is $\pi \rws$.
The parameters $j\ll{1,2}, e\uu 0$ are parameters
of the solution within our \emph{ansatz} but do not
correspond directly to quantities that would be conserved
for a generic trajectory with nonconstant $\r\ll i$.  However
they are proportional via $\r$-dependent constants to
the Noether charges that are conserved for a general trajectory.
The Noether charges are just the integrals of canonical
cojugates to Killing coordinates $X\uu 0 , \phi\ll i$.
\def\fsi{ \mshg{\rm(factor/sign?)} }
The canonical conjugate local variables are
\begin{align}
  \Pi\uu 0 &= \ffk 1 {2\pi\apr} \cc \ffk {e\uu 0} {\rws}\ , \\
  \Pi\ll{\phi\ll i} &= {1\over{2\pi \apr}} \lsqq \cc G\ll{\phi\ll i\phi\ll i} \phd\ll 1 + B\ll{\phi\ll i y\ll i} y\ll i\pr \cc \rsqq  %
  = \ffk {\r\ll i\sqd} {2\pi \apr\cc \rws \cc (1 +
    \e\ll i \sqd \r\ll i\sqd)} \lsqq \cc j\ll i + \ffk {\e\ll i L\ll
    i} {\pi} \cc \rsqq \ .
\end{align}
As usual N\oe ther's theorem gives
\bbb
\bp\uu\m =  \int\ll 0 \uu{\pi\rws}   \cc d\s\uu 1 \cc \Pi\uu\m\ .
\eee
Now we apply the e.o.m. and Virasoro constraints
within our \it ansatz \rm \rr{Ansatz}.  We see that 
our \it ansatz \rm satisfies the equations of motion
automatically, and the Viraroro constraints impose
a mass-shell condition. 
\def\cutoutA{
so given our \it ansatz \rm \rr{Ansatz}  we find
\bbb
\eng = \bp\uu 0 = {{e\uu 0}\over{2\apr}}\kko\llsk\llsk e\uu 0 = 2\apr \eng\ ,
\xxn
J\ll i  = \bp\ll{\phi\uu i} 
=  \ffk {\r\ll i\sqd} {2\apr\cc (1 + \e\ll i \sqd \cc \r\ll i \sqd)}
\cc  \lsqq \cc  j\ll i  +
\ffk {\e\ll i L\ll i} {\pi}   \cc \rsqq
\xxn
~~
\xxx
~~
\xxx
\kko\llsk\llsk j\ll i = 2\apr\cc J\ll i \cc \lrdd \cc \e\ll i \sqd + \ffk {1} {\r\ll i\sqd}\cc\rrdd -
 \ffk {\e\ll i L\ll i} {\pi}
\eee

With our \it ansatz \rm the (classical) Hamiltonian density \rr{CalHClassical}
becomes
\bbb
\mathcal{H} = \Pi\ll\m\dot{X}\uu\m - \mathcal{L}
-{1\over{4\pi}}\cc
(\pp\ll {\s\uu 0}\sqd 
+ \pp\ll{\s\uu 1}\sqd)\cc \Phi(X)
\xxn
= \Pi\ll\m\dot{X}\uu\m - \mathcal{L}
\eee
with
\bbb
\Pi\ll 0 \dot{X}\uu 0 = - \ffk {(e\uu 0)\sqd} {2\pi\apr\cc\rws\sqd} = - \ffk 
{2\apr \eng\sqd} {\pi \rws\sqd}
\xxn
\Pi\ll{\phi\uu i} \phd\uu i = \ffk {j\ll i} {\rws} \cdot \ffk {J\ll i} {\pi \rws} = 
\ffk {J\ll i} {\pi\rws\sqd} \cc\lsqq\cc 2\apr J\ll i \cc \lrdd \cc \e\ll i\sqd + \ffk {1} {\r\ll i\sqd} \cc \rrdd
-  \cc \ffk {\e\ll i L\ll i} \pi \cc\rsqq\kko
\xxn
G\ll{00} \dot{X}\ll 0 \sqd = -  \dot{X}\ll 0 \sqd = - \ffk {(e\ll 0)\sqd} {\rws\sqd} = - \ffk {4\apr\sqd 
\eng\sqd} {\rws\sqd}
\xxn
G\ll{\phi\uu i \phi\uu i} \phd\ll i\sqd = +G\ll{\phi\uu i \phi\uu i} \cc  \ffk {j\ll i \sqd} {\rws\sqd} 
\xxn
=  +G\ll{\phi\uu i \phi\uu i} \cc \ffk {1} {\rws\sqd}\cc
  \lsqq\cc 2\apr\cc J\ll i \cc \lrdd \cc \e\ll i \sqd + \ffk {1} {\r\ll i\sqd}\cc\rrdd -
 \ffk {\e\ll i L\ll i} {\pi} \cc \rsqq\sqd 
\xxn
= \ffk {\r\ll i \sqd} {1 + \e\ll i\sqd \r\ll i \sqd} \cc \ffk {1} {\rws\sqd}\cc
 \lsqq\cc 2\apr\cc J\ll i \cc \lrdd \cc \e\ll i \sqd + \ffk {1} {\r\ll i\sqd}\cc\rrdd -
\ffk {\e\ll i L\ll i} {\pi} \cc \rsqq\sqd 
\xxn
G\ll{y\uu i y\uu i}  y\uu i\pr{}\sqd = G\ll{y\uu i y\uu i} \cc \ffk { L\ll i \sqd }  { \pi\sqd \cc \rws\sqd }
\xxn = \ffk {1} {1 + \e\ll i \sqd \r\ll i\sqd} \cc \ffk { L\ll i \sqd }  { \pi\sqd \cc \rws\sqd }
\xxn
\e\uu{ab} B\ll{\m\n} X\uu\m\ll{,a} X\uu\n\ll{,b} = 2 B\ll{\m\n} \dot{X}\uu\m X\pr{}\uu\n
= 2 B\ll{\phi\uu i y\uu i} \phd\uu i y\uu i\pr
\xxn
\cl\ll{\rm L} =  \ffk {1} {4\pi\apr} \cc  \lsqq \cc - G\ll{\m\n}\dot{X}\uu\m \dot{X}\uu\n
+ G\ll{\m\n} X\pr{}\uu\m X\pr{}\uu\n + 2\cc B\ll{\m\n} \dot{X}\uu\m X\pr{}\uu\n \cc \rsqq 
\xxn
\ch = \ffk {1} {4\pi\apr} \cc \lsqq \cc + G\ll{\m\n}\dot{X}\uu\m \dot{X}\uu\n
+ G\ll{\m\n} X\pr{}\uu\m X\pr{}\uu\n \cc\rsqq
\eee
So the $\ch$-Virasoro constraint is
\bbb
(\dot{X}\uu 0){}\sqd = G\ll{\phi\uu i\phi\uu i} \cc \phd\uu i{}\sqd + G\ll{y\uu i y\uu i} y\uu i\pr{}\sqd
\xxn
= \ffk {4\apr\sqd \eng\sqd} {\rws\sqd} =\ffk{1}{\rws\sqd}\cc
 \sum\ll {i = 1}\uu 2 \cc \ffk {1} {1 + \e\ll i \sqd \cc\r\ll i\sqd}
\cc\lrdd \cc \r\ll i\sqd j\ll i \sqd(\r\ll i) + \ffk {L\ll i\sqd} {\pi\sqd} \cc\rrdd\ ,
\xxn
j\ll i(\r\ll i) \equiv 2\apr\cc J\ll i \cc \lrdd \cc \e\ll i \sqd + \ffk {1} {\r\ll i\sqd}\cc\rrdd -
 \ffk {\e\ll i L\ll i} {\pi}
\eee
which gives
\bbb
 \eng\sqd = \ffk{1}{4\apr\sqd}\cc
 \sum\ll {i = 1}\uu 2 \cc \ffk {1} {1 + \e\ll i \sqd \cc\r\ll i\sqd}
\cc\lrdd \cc \r\ll i\sqd j\ll i \sqd(\r\ll i) + \ffk {L\ll i\sqd} {\pi\sqd} \cc\rrdd\ .
\eee
The $\mathcal{P}$-Virasoro constraint is satisfied automatically, given our \it ansatz. \rm

}

In the case where $J\ll 1 =  L\ll 1 = 0$, the mass formula is particularly
transparent, so let us consider that case.  The definitions of
the conjugate momenta give
\bbb
j\ll 1(\rho) = 0\ ,
\xxn
j\ll 2 (\rho) = 2\apr J\ll 2 \cc \lrdd \cc \e\ll 2 \sqd + \ffk {1} {\r\ll 2\sqd} \cc\rrdd - \ffk
{\e\ll 2 L\ll 2} {\pi} \kko\ 
\eee
and the mass-shell condition imposed by the classical 
Virasoro constraint is
\bbb
\label{eq:string-state-energy}
\eng\sqd = \lrdd \cc \e\ll 2 J\ll 2 -   \ffk {L\ll 2} {2\pi\apr} \cc \rrdd\sqd + 
 \ffk {J\ll 2\sqd} {\r\ll 2\sqd} 
\cc \ppo
\eee
Minimizing $\eng\sqd$ with respect to $\r\ll 2$ gives
\bbb
\eng\ll{\rm BPS} = \left |\cc \e\ll 2 J\ll 2 -   \ffk {L\ll 2} {2\pi\apr} \cc \right | \ppo
\eee

\subsection{String-theoretic description of the BPS states the reciprocal frame}

In this section we want to give a unified string theoretical description of the \oscs and \dozzs as they appear in the reciprocal frame described in Equation~\eqref{eq:reciprocal-frame}. First let us follow them along the change of frame, as in Table~\ref{tab:Everything}.
\begin{itemize}
\item An \osc is a \D0--brane in the fluxtrap, localized at the origin. In the lift to M--theory this turns into a momentum mode in the direction \( x_{10} \), which is also the way in which it appears in the reciprocal frame.
\item A \dozz is a fundamental string extended in \( x_9 \) with a momentum in \( \phi_2 \). This is lifted to an \M2--brane with momentum in \( \phi_2 \) and eventually, in the reciprocal frame, it turns into a \D3--brane in \( (x_6, x_9, x_{10}, \tilde \sigma_2) \) with an electric flux in the \( \tilde \sigma_2 \) direction. 
\end{itemize}
The main feature of this last frame is that we now have only \emph{one} kind of \ac{bps} state carrying \emph{both} types of charge and living at a finite radius $\bar\rho_2$, as we will show in the following. 

\begin{figure}
  \centering
  \includegraphics[width=.6\textwidth]{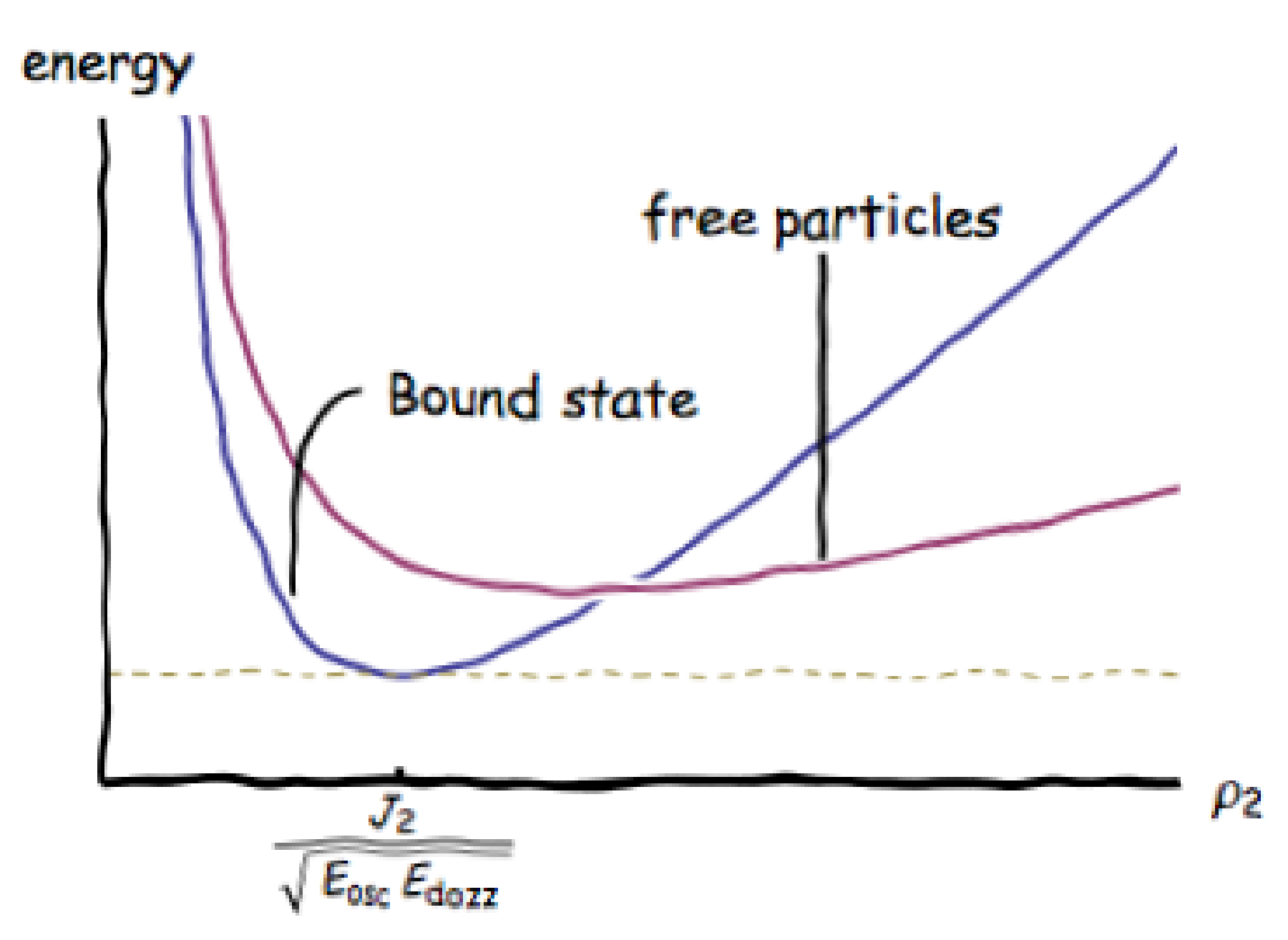}
  \caption{A qualitative plot of the 
  energy of the bound state and of two coincident free particles as function of \( \rho_2 \) for \( \rho_1 = 0 \).  }
   \label{fig:Binding-Energy}
\end{figure}

A unified description is possible if we introduce a \D3--brane extended in \( (x_6, x_9, x_{10}, \tilde \sigma_2) \) with an electric field in \( \tilde \sigma_2 \), velocity \( v \) in \( x_{10} \) and another component of the electric field in \( x_{10} \), which is required by the coupling of \( v \) in the \ac{dbi} action (see the cartoon in Figure~\ref{fig:brane-cartoon}). 
Consider the embedding
\begin{equation}
  \begin{aligned}
    x_6 &= \im \zeta_0 \  , & x_9 &=\frac{L_2}{\pi \rws} \zeta_1 \  , & \tilde \sigma_2 &= 2 \pi \alpha' \epsilon_2\frac{\zeta_2}{\kappa} \  , & x_{10} &= \zeta_3 + v \zeta_0 \  , & x_8 &= 0 \  , \\
    \rho_1 &= \text{const.} \  , & \rho_2 &= \text{const.} \  , & \rho_3 &= 0 \  , & x_7 &= 0 \  ,
  \end{aligned}
\end{equation}
with a \( U(1) \) gauge field with the following components turned on:
\begin{align}
  F_{02} &= \frac{1}{\kappa} f_{02}\,, & F_{03} &= \frac{1}{\kappa} f_{03} \  .  
\end{align}
The corresponding \ac{dbi} action reads:
\begin{equation}
  \begin{aligned}
    S &= - \mu_3 \int \di^4 \zeta \eu^{-\Phi} \sqrt{ - \det (g + B + 2 \pi \alpha' F) } + \mu_3 \int \exp ( B + 2 \pi \alpha' F ) \wedge \sum_q C_q \\
    &= \frac{\epsilon_1 L_2 R_{10}}{2 \pi \alpha'} \frac{ \epsilon_2 \rho_2^2 \left( f_{02} + v f_{03} \right) - \sqrt{ \left( 1 + \epsilon_2^2 \rho_2^2 \right) \left( 1 + \rho_2^2 f_{03}^2 \left( 1 + \epsilon_1^2 \rho_1^2 \right)  \right) - \rho_2^2 \left(f_{02} + v f_{03} \right)^2 } }{1+ \epsilon_2^2 \rho_2^2} \  .
  \end{aligned}
\end{equation}
In order to evaluate the energy we pass to the Hamiltonian formalism. There are three conjugate momenta, corresponding to the components of the electric field and the velocity, \emph{viz.}:
\begin{align}
  J_2 &= \frac{\delta S}{\delta f_{02}} \ , &   D_3 &= \frac{\delta S}{\delta f_{03}} \  , 
 &  P &= \frac{\delta S}{\delta v} \  .
\end{align}
First we observe that since the action only depends on \( f_{02} \) and \( v \) via  \( (f_{02} + v f_{03}) \), the momenta satisfy
\begin{equation}
  P = f_{03} J_2 \  .  
\end{equation}
Using this fact we can invert the relations and find that
\newcommand*{\SumOfSquares}{\ensuremath{{\left(\frac{\epsilon_1 R_{10}}{2 \pi \alpha'} L_2 - \epsilon_2 J_2\right)}^2 + {\left(\frac{J_2}{\rho_2} \right)}^2}}
\newcommand*{\OnePlusSquare}{\ensuremath{1 + \rho_2^2 \left(1 + \epsilon_1^2 \rho_1^2 \right) {\left(\frac{P}{J_2} \right)}^2}}
\begin{equation}
  \begin{aligned}
    v ={}& \textstyle \frac{\rho_2^2}{J_2^2} \left( 1 + \epsilon_1^2 \rho_1^2 \right) \sqrt{\frac{\SumOfSquares}{\OnePlusSquare}} P + \frac{D_3}{J_2} \ ,\\
    f_{02} ={}& \textstyle \epsilon_2 \left( - \frac{\epsilon_1 L_2 R_{10}}{2 \pi \alpha'} + \epsilon_2 J_2 + \frac{J_2}{\rho_2^2 \epsilon_2} \right) \sqrt{ \frac{\OnePlusSquare}{\SumOfSquares}} \\
    & \textstyle - \frac{\rho_2^2}{J_2} \left( 1 + \epsilon_1^2 \rho_1^2 \right) \sqrt{\frac{\SumOfSquares}{\OnePlusSquare} } {\left( \frac{P}{J_2} \right)}^2 + \frac{P D_3}{J_2^2}\ , \\
    f_{03} ={}& \textstyle \frac{P}{J_2}\ ,
  \end{aligned}
\end{equation}
and derive the Hamiltonian:
\begin{equation}
  \begin{aligned}
    H &= f_{02} J_2 + f_{03} D_3 + P v - S \\ 
    &= \frac{D_3 P}{J_2} + \sqrt{ \left[ \SumOfSquares \right] \left[ \OnePlusSquare \right] }\ .
  \end{aligned}
\end{equation}
The novelty of this frame is the contemporary presence of all the momenta for a single \D3--brane.
First consider the electric displacement \( D_3 \). In the reciprocal frame that we are using now, the electric field \( F_{03} \)  is equivalently described by a fundamental string winding around \( x_{10} \) and dissolved on the \D3--brane. In turn, in the M--theory frame, this corresponds to an \M2--brane winding around \( \sigma_1 \) and \( x_{10} \). Finally, in the fluxtrap this is again a fundamental string winding around \( \phi_1 \). At this point it is clear that \( D_3 \) cannot correspond to a conserved \ac{bps} charge, since the fundamental string is wrapped around a contractible cycle. 
Hence, in order to find the \ac{bps} condition we should set \( D_3 = 0 \) and maximize the energy with respect to the positions \( \rho_1 \) and \( \rho_2 \) for fixed values of the momenta \( J_2 \) and \( P \).
Before doing that, it is instructive to consider the limits of vanishing momenta and compare with the results of the previous section.
\begin{itemize}
\item For \( P = 0 \) the energy is 
  \begin{equation}
    \eng_{P=0} = \sqrt{ \SumOfSquares } \  ,
  \end{equation}
  which coincides with the one that had been obtained for the string states in the fluxtrap in Equation~\eqref{eq:string-state-energy} once we map the values of \( \alpha' \) in the two frames using Equation~\eqref{eq:length-mapping}:
  \begin{equation}
    {\left. \frac{1}{\gAO {(\alpha')}^{3/2}} \right|}_{\Omega} = {\left. \frac{\epsilon_1 }{ \alpha'} \right|}_{\text{rec}} \ .
  \end{equation}
\item For \( L_2 = J_2 = 0 \) the energy is
  \begin{equation}
    \eng_{L_2 = J_2 = 0} = P \sqrt{ \left( 1 + \epsilon_1^2 \rho_1^2  \right) \left( 1 + \epsilon_2^2 \rho_2^2 \right) } \  ,
  \end{equation}
  which coincides with the energy of \( n_{\D0} \) \D0 branes in the fluxtrap found in Equation~\eqref{eq:D0-energy} if we observe that \( P \) is the quantized momentum around a circle of radius \( R_{10} \):  
  \begin{equation}
    P = \frac{n_{\D0}}{R_{10}} = \frac{n_{D0}}{\gAO \lO} \  ,   
  \end{equation}
  where we used the definition of \( R_{10} \)  in Equation~\eqref{eq:gIIAOmega-R10}.
\end{itemize}
Let us now proceed to the minimization. Consider the square of the energy,
\begin{equation}
  \eng^2 = \left[ \SumOfSquares \right] \left[ \OnePlusSquare \right].
\end{equation}
We want to minimize with respect to the two radii. For \( \rho_1 \) we find that
\begin{equation}
  \frac{\del \eng^2}{\del \rho_1^2} = \epsilon_1^2 \rho_1 \rho_2^2 \left( \frac{P}{J_2} \right)^2 \left(\SumOfSquares \right) = 0 \  ,
\end{equation}
which is satisfied for \( \rho_1 = 0 \) or for \( P = 0 \). Keeping \( \rho_1 = 0 \), we minimize with respect to \( \rho_2^2 \) and find
\begin{equation}
  \frac{\del \eng^2}{\del \rho_2^2} = - \frac{J_2^2}{\rho_2^4} + \frac{P^2}{J_2^2} \left(\frac{\epsilon_1 R_{10}}{2 \pi \alpha'} L_2- \epsilon_2 J_2 \right)^2 = 0 \  ,
\end{equation}
which is satisfied for
\begin{equation}
\boxed{ \rho_2^2 = \bar \rho_2^2 = \frac{J_2^2}{P \left( \frac{\epsilon_1 R_{10}}{2 \pi \alpha'} L_2 - \epsilon_2 J_2 \right)} \,. }
\end{equation}
Putting this back into the expression for the energy we find that the energy of the \ac{bps} states is obtained as linear combination of the momenta:
\begin{equation}
\boxed{  \eng_{\ac{bps}} = \abs{ \frac{\epsilon_1 R_{10}}{2 \pi \alpha'} L_2 - \epsilon_2 J_2 + P } \  .}
\end{equation}
The two obvious limits are:
\begin{itemize}
\item for \(  L_2 = J_2 = 0\) we recover the \osc with energy \( \eng_{\text{osc}} = P \) that lives at the origin \( \bar \rho_2 = 0 \);
\item for \( P = 0 \) we recover the \dozz with energy \( \eng_{\text{\textsc{dozz}}} =  \abs{ \frac{\epsilon_1 R_{10}}{2\pi \alpha'}L_2 - \epsilon_2 J_2} \) that lives at infinity \( \bar \rho_2 \to \infty \). 
\end{itemize}
When both charges are turned on, the \D3--brane lives at finite radius, given by
\begin{equation}
  \bar \rho_2 = \frac{J_2}{\sqrt{\eng_{\text{osc}} \eng_{\text{\textsc{dozz}}}}} \ .
\end{equation}
The fact that the energies sum linearly as \( \eng_{\ac{bps}} = \eng_{\textsc{osc}} + \eng_{\textsc{dozz}} \)  is somewhat surprising. It means that the states are only marginally stable with respect to their decay into separate  \oscs and \dozzs. On the other hand, the fact that the bound state lives at a precise value of the radius \( \bar \rho_2 \), while the components are localized at different places (\( \rho_2 = 0  \) and \( \rho_2 \to \infty \)) means that any decay would have to tunnel over a barrier, since a state that is broken apart locally would have an energy that is strictly larger than the one of the bound state (see Figure~\ref{fig:Binding-Energy}). We therefore expect that any such decay process, should it happen, would have to be nonperturbatively suppressed. It would be interesting to study the stability of the \ac{bps} states from the point of view of the six-dimensional \( (2,0) \) theory, where it may possibly be understood in terms of wall-crossing between the two limits represented by the compactifications to four dimensions, but this goes beyond the scope of this note.

\section{Conclusions}
\label{sec:conclusions}

In this note we have studied different limits of the dynamics of a pair of M5--branes in string- and M--theory.  The two resulting four-dimensional theories are the \( \Omega \)--deformation of \( \mathcal{N} = 4 \)  super Yang--Mills on \( \setR^4 \) and a new supersymmetric non-Lorentz-invariant theory  on \( \setR_+^2 \times T^2 \) described by the \emph{reciprocal action}. These theories are obtained in terms of effective theories on the \D3--branes resulting from different limits of \M5--branes in the M--theory fluxtrap introduced by the authors in~\cite{Hellerman:2012zf}. We have identified the \ac{bps} states appearing on the \ac{sym} side of the \ac{agt} correspondence in terms of fundamental strings and \D0--branes in the \tIIA fluxtrap, and followed them through the duality chain to the reciprocal frame where they are identified with \D3--branes carrying both electric fields and momentum in the $x_{10}$ direction, localized at a finite value of the radial direction \( \rho_2 \) depending on the charges. The fact that we are dealing with two distinct types of states localized in different positions (the origin and infinity) in one duality frame, and states carrying both charges localized in one place in the other duality frame suggests the presence of new phenomena which are only accessible via a microscopic description such as the one proposed in this article.

One of the main points we would like to stress is that the reciprocal theory exhibits some characteristic similarities to the Liouville field theory in the \ac{agt} correspondence: its loop-counting parameter is \( b^2 = \epsilon_2 / \epsilon_1 \) and S--duality is realized as the exchange \( b \leftrightarrow 1/b \). The reciprocal theory is however intrinsically four-dimensional. This is because the original six-dimensional theory lives on \( \setR^4_\Omega \times T^2 \), where \( \setR^4_\Omega \) is the deformation of \( \setR^4 \) into the product of two cigars with asymptotic radii \( 1/\epsilon_1 \) and \( 1/ \epsilon_2 \). As a result, only two compact directions are available for the reduction and the reciprocal theory therefore lives on \( \setR_+^2 \times T^2 \). 
In order to construct the true Liouville field theory as a compactification of an \M5--brane in an eleven-dimensional fluxtrap background, it would be necessary to start from a geometry of the type \( S^4_\Omega \times \Sigma \) in order to be able to reduce on the four-dimensional part and realize a two-dimensional theory on the Riemann surface \( \Sigma \). This topic will be addressed in a future article.

\subsection*{Acknowledgements}
 
It is our pleasure to thank Wolfgang Lerche for illuminating discussions.
The work of S.H. was supported by the World Premier International Research Center Initiative, \textsc{mext}, Japan, and also by a Grant-in-Aid for Scientific Research (22740153) from the Japan Society for Promotion of Science (\textsc{jsps}).

\appendix

\section{Supersymmetry in the bulk}
\label{sec:supersymmetry}

In this section we follow the duality chain to obtain the explicit expressions for the Killing spinors in each frame.
\newcommand*{\scL}{\textsc{l}}
\newcommand*{\scR}{\textsc{r}}

\begin{itemize} 
\item The initial flat space, before imposing the Melvin identifications, has \( 32 \) supercharges. The corresponding \tIIA Killing spinors can be put in the form 
  \begin{equation}
    \begin{cases}
      k_{\scL} = \exp[ \frac{\theta_1}{2} \gamma_{01} + \frac{\theta_2}{2} \gamma_{23} + \frac{\theta_3}{2} \gamma_{45} ] \eta_{\scL} \  , \\
      k_{\scR} = \exp[ \frac{\theta_1}{2} \gamma_{01} + \frac{\theta_2}{2} \gamma_{23} + \frac{\theta_3}{2} \gamma_{45} ] \eta_{\scR}  \  ,
    \end{cases}
  \end{equation}
  where \( \eta_{\substack{\scL \\ \scR}} \) are constant spinors satisfying
  \begin{equation}
    \gamma_{11} \eta_{\substack{\scL \\ \scR}} = \pm \eta_{\substack{\scL \\ \scR}} \  .
  \end{equation}
\item Imposing the Melvin identifications and passing to the disentangled variables \( \phi_i \), the exponential becomes
  \begin{multline}
    \exp[ \frac{\theta_1}{2} \gamma_{01} + \frac{\theta_2}{2} \gamma_{23} + \frac{\theta_3}{2} \gamma_{45} ] \\ 
    = \exp[ \frac{\phi_1}{2} \gamma_{01} + \frac{\phi_2}{2} \gamma_{23} + \frac{\phi_3}{2} \gamma_{45} ] \exp[ \epsilon_1 \frac{\wt R_1 \tilde u_1}{2} \left( \gamma_{45} - \gamma_{01} \right) + \epsilon_2 \frac{\wt R_2 \tilde u_2}{2} \left( \gamma_{45} - \gamma_{23} \right) ] .
  \end{multline}
  The \( \epsilon \)--dependent terms are not invariant under the period \( \tilde u_i \mapsto \tilde u_i + 2 \pi \) and have to be projected out using
  \begin{align}
    \Pi_1 &= \tfrac{1}{2}\left( \gamma_{45} - \gamma_{01} \right) \,, &  \Pi_2 &= \tfrac{1}{2}\left( \gamma_{45} - \gamma_{23} \right) \  .
  \end{align}
  Each projector breaks one half of the supersymmetry. The remaining eight Killing spinors are: 
  \begin{equation}
    \begin{cases}
      k_{\scL} = \Pi_1 \Pi_2 \exp[ \frac{\phi_1}{2} \gamma_{01} + \frac{\phi_2}{2} \gamma_{23} + \frac{\phi_3}{2} \gamma_{45} ] \eta_{\scL} \  , \\
      k_{\scR} = \Pi_1 \Pi_2 \exp[ \frac{\phi_1}{2} \gamma_{01} + \frac{\phi_2}{2} \gamma_{23} + \frac{\phi_3}{2} \gamma_{45} ] \eta_{\scR}  \  .
    \end{cases}
  \end{equation}
\item The two T--dualities in \( \tilde u_1 \) and \( \tilde u_2 \) leave the left-moving spinors invariant and transform the right-moving ones,
  \begin{equation}
    \begin{cases}
      k_{\scL} = \Pi_1 \Pi_2 \exp[ \frac{\phi_1}{2} \gamma_{01} + \frac{\phi_2}{2} \gamma_{23} + \frac{\phi_3}{2} \gamma_{45} ] \eta_{\scL} \  , \\
      k_{\scR} = \Gamma_{u_1} \Gamma_{u_2} \Pi_1 \Pi_2 \exp[ \frac{\phi_1}{2} \gamma_{01} + \frac{\phi_2}{2} \gamma_{23} + \frac{\phi_3}{2} \gamma_{45} ] \eta_{\scR}  \  ,
    \end{cases}
  \end{equation}
  where \( \Gamma_{u_i} \) are the gamma matrices in the direction \( u_i \). 
\item Introducing the angle variable \( \psi = \phi_1 + \phi_2 + \phi_3 \) transforms the exponential into
  \begin{equation}
     \exp[ \frac{\phi_1}{2} \gamma_{01} + \frac{\phi_2}{2} \gamma_{23} + \frac{\phi_3}{2} \gamma_{45} ] = \exp[ \frac{\phi_1}{2} \left( \gamma_{01} - \gamma_{45} \right) + \frac{\phi_2}{2} \left( \gamma_{23} - \gamma_{45} \right) + \frac{\psi}{2} \gamma_{45} ]\  ,
  \end{equation}
  the dependence of \( \phi_1 \) and \( \phi_2 \)  is projected out by \( \Pi_{1,2} \) and the Killing spinors read
  \begin{equation}
    \begin{cases}
      k_{\scL} = \Pi_1 \Pi_2 \exp[ \frac{\psi}{2} \gamma_{45} ] \eta_{\scL} \  , \\
      k_{\scR} = \Pi_1 \Pi_2 \Gamma_{u_1} \Gamma_{u_2} \exp[ \frac{\psi}{2} \gamma_{45} ] \eta_{\scR}  \  .
    \end{cases}
  \end{equation}
  The fact that the spinors do not depend on \( \phi_1 \) or \( \phi_2 \) is the reason why the following changes of frame do not break local supersymmetries.
\item The lift to M--theory is obtained by multiplying the spinors by a conformal factor that depends on the \tIIA dilaton,
  \begin{equation}
    k_{M} = \eu^{- \Phi/6} k_{IIA}    ,
  \end{equation}
  where the dilaton is the one in the fluxtrap of Equation~\eqref{eq:omega-frame}:
  \begin{equation}
    \eu^{-\Phi/ 6} = {\left[ \left( 1 + \epsilon_1^2 \rho_1^2 \right) \left( 1 + \epsilon_2^2 \rho_2^2 \right) + \rho_3^2 \left( \epsilon_1^2 \Delta_2^2 + \epsilon_2^2 \Delta_1^2 \right) \right]}^{1/12}\ .
  \end{equation}
\item The reduction to \tIIA is obtained in the same way, this time using the dilaton in the reciprocal frame of Equation~\eqref{eq:reciprocal-frame}. The overall result is that
  \begin{equation}
    \begin{cases}
      k_{\scL} = H(\rho_1, \rho_2) \Pi_1 \Pi_2 \exp[ \frac{\psi}{2} \Gamma_{45} ] \eta_{\scL} \  , \\
      k_{\scR} = H(\rho_1, \rho_2) \Gamma_{u_1} \Gamma_{u_2} \Pi_1 \Pi_2 \exp[ \frac{\psi}{2} \Gamma_{45} ] \eta_{\scR}  \  ,
    \end{cases}
  \end{equation}
  where
  \begin{equation}
    H(\rho_1, \rho_2) = \left[ \rho_1^2 \left( 1 + \epsilon_2^2 \rho_2^2 \right) + \rho_3^2 \left( 1 + \epsilon_2^2 \left( \rho_1^2 + \rho_2^2 \right) \right) \right]^{1/8} \  .
  \end{equation}
\item The final T--duality to \tIIB changes the right-moving spinor and leads us to the final expression for the eight Killing spinors preserved in the reciprocal frame:
  \begin{equation}
    \begin{cases}
      k_{\scL} = H(\rho_1, \rho_2) \Pi_1 \Pi_2 \exp[ \frac{\psi}{2} \Gamma_{45} ] \eta_{\scL} \  , \\
      k_{\scR} = H(\rho_1, \rho_2) \Gamma_\phi \Gamma_{u_1} \Gamma_{u_2} \Pi_1 \Pi_2 \exp[ \frac{\psi}{2} \Gamma_{45} ] \eta_{\scR}  \  ,
    \end{cases}
  \end{equation}
  where \( \Gamma_\phi \) is the gamma matrix in the \( \phi_2 \) direction.
\end{itemize}

\printbibliography

\end{document}